\documentclass{aa}  

\usepackage{graphicx}
\usepackage{color}
\usepackage{soul}

\usepackage{txfonts}

\begin{document}

   \title{Chemical composition of the young massive cluster NGC 1569-B}
 \author{A. Gvozdenko\inst{1},
          S.\ S.\ Larsen\inst{1},
          M. A. Beasley\inst{2,3}
          \and
          J. Brodie\inst{4,5}}

   \institute{Department of Astrophysics/IMAPP, Radboud University, PO Box 9010,
6500 GL, The Netherlands
         \and
             Instituto de Astrofísica de Canarias, Calle Vía Láctea, E-38206 La Laguna, Spain
        \and    
            Departamento de Astrofísica, Universidad de La Laguna, E-38206 La Laguna, Spain
        \and     
             Centre for Astrophysics and Supercomputing, Swinburne University, John Street, Hawthorn VIC 3122, Australia
        \and     
              University of California Observatories, 1156 High Street, Santa Cruz, CA 95064, USA  
             }

   \date{Accepted 9 Aug 2022}

  \abstract
   {The chemical composition of young massive clusters (YMCs) provides stellar population information on their host galaxy. As potential precursors of globular clusters (GCs), their properties can help us understand the origins of GCs and their evolution.}
   {We present a detailed chemical abundance analysis of the YMC NGC 1569-B. The host galaxy, NGC~1569, is a dwarf irregular starburst galaxy at a distance of 3.36$\pm$0.20 Mpc. We derive the abundances of the $\alpha$, Fe-peak, and heavy elements.}
   {We determined the abundance ratios from the analysis of an optical integrated-light (IL) spectrum of NGC 1569-B, obtained with the HIRES echelle spectrograph on the Keck I telescope.
  We considered different red-to-blue supergiant ratios ($N_{RSG}/N_{BSG}$), namely: the ratio obtained from a theoretical isochrone ($N_{RSG}/N_{BSG}$=1.24), the ratio obtained from a resolved colour-magnitude diagram of the YMC ($N_{RSG}/N_{BSG}$=1.53), and the ratio that minimises the
  $\chi^2$ when comparing our model spectra with the observations ($N_{RSG}/N_{BSG}$=1.90). We adopted the latter ratio for our resulting chemical abundances.}
   {The derived iron abundance is sub-solar with [Fe/H] = $-0.74\pm0.05$. 
   In relation to the  scaled solar composition, we find enhanced $\alpha$-element abundances, $\text{[<Mg,Si,Ca,Ti>/Fe]}=+0.25\pm$0.11, with a particularly high Ti abundance of +0.49$\pm$0.05. Other super-solar elements include $\text{[Cr/Fe]}=+0.50\pm$0.11, $\text{[Sc/Fe]}=+0.78\pm$0.20, and $\text{[Ba/Fe]}=+1.28\pm$0.14, while other Fe-peak elements are close to scaled solar abundances: ($\text{[Mn/Fe]}=-0.22\pm$0.12 and $\text{[Ni/Fe]}=+0.13\pm$0.11).}
   {The composition of NGC 1569-B resembles the stellar populations of the YMC NGC 1705-1, located in a blue compact dwarf galaxy. The two YMCs agree with regard to $\alpha$-elements and the majority of the Fe-peak elements, except for Sc and Ba, which are extremely super-solar in NGC~1569-B -- and higher than in any YMC studied so far. The blue part of the optical spectrum of a young population is still a very challenging wavelength region to analyse using IL spectroscopic studies. This is due to the uncertain contribution to the light from blue supergiant stars, which can be difficult to disentangle from turn-off stars, even when resolved photometry is available. 
   We suggest that the comparison of model fits at different wavelengths offers a route to determining the red-to-blue supergiant ratio from IL spectroscopy.}

   \titlerunning{Chemical composition of the young massive cluster NGC 1569-B}
   \authorrunning{A. Gvozdenko et al.}
   \maketitle

\section{Introduction}

The element abundances of star clusters can reveal information on their star formation (SF) history and the evolution of their host galaxies. Indeed, obtaining accurate abundances for star clusters
is a crucial goal for both galactic and extragalactic targets, since these parameters facilitate studies of the evolution of both stars and galaxies. 

In recent decades, the availability of 8-10 m class telescopes has made it possible to study the spectra coming from stars outside of the Milky Way (MW), including those located in 
neighbouring galaxies such as the Small and Large Magellanic Clouds (SMC and LMC) as well as nearby dwarf galaxies \citep{Hill1999}. It has been found that for old stellar populations with low iron abundances ([Fe/H]), the $\alpha$-abundances of stars in dwarf galaxies exhibit similar patterns to those characterising the MW, with enhanced abundances (relative to scaled-solar composition).
However, as [Fe/H] increases, the [$\alpha$/Fe] ratio in dwarfs reaches lower values than in the MW \citep{Tolstoy2009}. 

Beyond the Local Group, detailed information on the chemical composition of stellar populations is much more scarce, especially with regard to detailed abundance patterns. Based on observations of \ion{H}{ii} regions, abundances of a limited set of elements (such as oxygen and nitrogen) can be inferred in galaxies with ongoing or recent SF \citep{Shields1990}, and the spectroscopy of individual supergiant (SG) stars has been used to measure overall metallicities \citep{Larsen2008, Davies2013}. However, differences in the elements typically included in these types of analyses have made it difficult to compare these results directly with the data for old stellar populations in nearby galaxies.

The chemical composition of young massive clusters (YMCs) provides valuable information on stellar population.
These objects allow us to gain insight into SF under more extreme conditions than those encountered in nearby star-forming regions, and they are of particular interest, particularly since an ongoing debate in the literature has questioned whether these objects might be the progenitors of the old globular clusters (GCs) that we see today \citep{Lardo2017, Cabrera-Ziri2020}.
Additionally, it is interesting to compare the chemical abundance patterns that we see in our Local Group with systems outside of it. Thus, it is important to study extragalactic clusters to explore their chemical composition, and YMCs can serve as ideal targets thanks to their intrinsic high luminosities. 

The high light-to-mass ratios of young clusters, such as YMCs, allow us to study their abundances using integrated-light (IL) spectroscopy. The IL spectra for the majority of star clusters are broadened by only up to 10 km s$^{-1}$ which implies that relatively high spectral resolutions ($\lambda/\Delta\lambda \ga$ 10,000) are accessible; thus, detailed chemical abundances can be extracted from their IL spectra. In general, this is done by modelling the IL spectra using information from the Hertzsprung–Russell diagram (HRD), combined with atmospheric models and synthetic stellar spectra \citep{Davies2010, Larsen2012, Sakari2021}. One method that uses this technique was introduced and described in \cite{Larsen2012}, and it has been used to derive abundances for iron and $\alpha$-elements, as well as light- and n-capture elements for old GCs in Local Group dwarf and spiral galaxies \citep{Larsen2014, Larsen2022}. 
Extragalactic YMCs studied with this method include the clusters NGC 1313-379 (in the barred spiral NGC~1313) and NGC 1705-1, in a blue compact dwarf \citep{Hernandez2017}. The derived metallicities were: $\text{[Fe/H]}=-0.84\pm$0.07 and $\text{[Fe/H]}=-0.78\pm$0.10 for NGC 1313-379 and NGC 1705-1, respectively; while the $\alpha$ abundance is solar like for NGC 1313-379 with [$\alpha$/Fe] = +0.06$\pm$0.11 and a super-solar [$\alpha$/Fe] = +0.32$\pm$ 0.12 for NGC 1705-1.

It is important to increase the number of YMCs for which the chemical composition has been measured in detail. 
In \cite{Larsen2006}, iron and $\alpha$ abundances were derived for a young super-star cluster SSC in the nearby spiral galaxy NGC 6946 using near-infrared (NIR) observations from the Near Infrared Spectrograph (NIRSPEC) instrument on the Keck~II telescope. These authors found a super-solar [$\alpha$/Fe] and sub-solar metallicity of $\text{[Fe/H]}=-0.45\pm$0.08. 
Other studies of YMCs include \citet{Lardo2017} and \citet{Cabrera-Ziri2016} where YMCs were studied to search for GC-like abundance patterns.

Such patterns can be used to infer star-to-star variations of some light elements (e.g. Na-O anticorrelation), which are linked to the presence of `multiple populations' (MPs) in GCs \citep{Bastian2018}. In particular, it is interesting to try and identify the first signs of the MPs within the YMCs that would lead to the patterns seen in the GCs. However, no clear indications of the abundance patterns associated with MPs have yet been found in any clusters younger than about 2~Gyr \citep{Martocchia2018, Martocchia2019}.

\begin{figure}
\centering
\includegraphics[width=9cm]{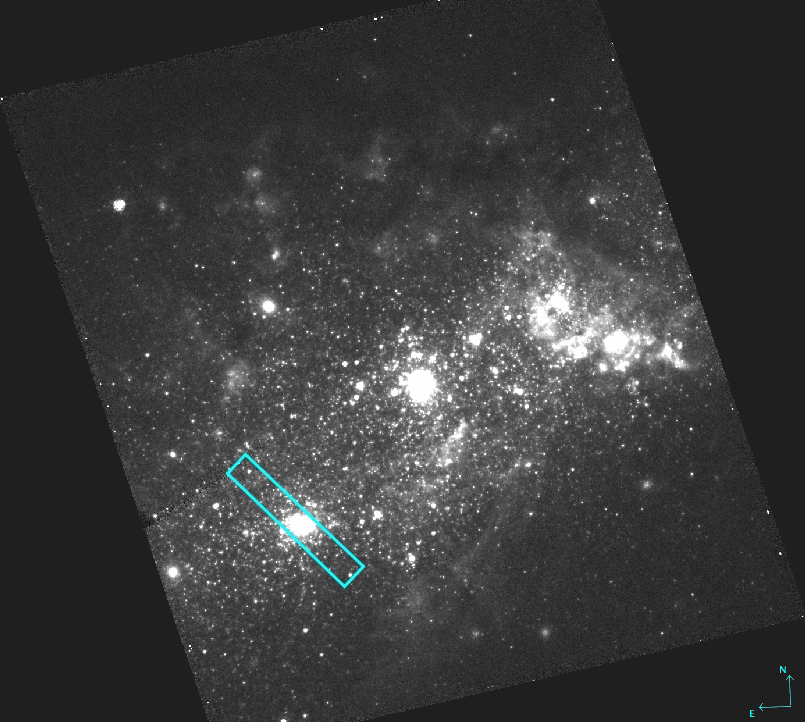}
\caption{Image of the NGC 1569 using F555W filter from HST ACS. The slit is located on top of the NGC 1569-B, indicating the region the spectra were obtained from.} 
\label{target_img}%
\end{figure}

In this study, we derive the detailed chemical abundances of a YMC NGC 1569-B. It is located in a dwarf irregular galaxy (dIrr) NGC 1569, which has two bright YMCs: NGC 1569-A (a binary cluster) and NGC 1569-B. The masses were estimated to be $\approx10^6$\(M_\odot\) for the binary cluster and $(4.4\pm1.1)\times10^5$ \(M_\odot\) for NGC 1569-B \citep{Ho1996, Sternberg1998, Larsen2008}. Both of these clusters are more massive than any of the young clusters in the MW or LMC \citep{Anders2004}. The cluster investigated in this work is the older and less bright one out of the two YMCs. Figure\,\ref{target_img} shows a \textit{Hubble} Space Telescope (HST) image of the target with the slit used for our spectroscopic observations. NGC~1569-B is particularly interesting given its mass, since such clusters are rare in the Milky Way and even in the Local Group. The massive clusters can provide clues about the MPs seen in old GCs.

\begin{table}
\caption{Basic parameters of NGC 1569.}           
\label{galaxy_param} 
\centering 
\begin{tabular}{l c l}   
\hline
\hline   
Quantity & Observed value & References \\ 
\hline                
   D & 2.2 $\pm$ 0.6 Mpc & 1 \\  
     & 3.36 $\pm$ 0.20 Mpc & 2 \\
   d & 1.85 kpc & 3 \\
   M$_{gas}$ & (1.5 $\pm$ 0.3) $\times$ 10$^8$ \(M\odot\) & 1 \\
   M$_{tot}$ & 3.3 $\times$ 10$^8$ \(M\odot\) & 1 \\
   Z & 0.004 & 4  \\
   SFR & $\sim$ 0.5 \(M\odot\) yr$^{-1}$& 5   \\
    & 0.32 \(M\odot\) yr$^{-1}$& 6   \\
\hline                         
\end{tabular}
\tablefoot{D - the distance to the galaxy; d - maximum optical size; M$_{gas}$ - total atomic gas mass (H I + He); M$_{tot}$ - total galaxy mass.}
\tablebib{(1)~\cite{Israel1988}, (2)~\cite{Grocholski2008}, (3)~\cite{Stil_Israel2002}, (4)~\cite{Gonzales1997}, (5)~\cite{Greggio1998},  (6)~\cite{Hunter2004}.}
\end{table}

Table\,\ref{galaxy_param} lists the basic parameters of NGC 1569. The stellar population in this dIrr is dominated by stars that are younger than a few tens of Myrs, while some old stars are also present \citep{Greggio1998}. \cite{Anders2004} confirmed a major burst 25 Myr ago in NGC 1569. During this time, a large number of clusters ($\approx$150) were formed, including the massive clusters NGC 1569-A and NGC 1569-B. 
Evidence of older episodes of SF from 1.5 Gyr to 150 Myr ago has been presented by \cite{Vallenari1996}. NGC1569 is found to have an age gradient, where the youngest stars are mainly concentrated in the super star clusters (NGC 1569-A and NGC 1569-B), intermediate-age stars are more uniformly distributed, and the oldest stars are located in the outskirts of the star bursting regions \citep{Aloisi2001}.

The stellar content of the B cluster (together with some abundances) were previously investigated with NIRSPEC in \cite{Larsen2008}. Those derived abundances are: Fe abundance of $\text{[Fe/H]}=-0.63 \pm 0.08$, $\alpha$-abundance [$\alpha\text{/Fe]}=+0.31 \pm 0.09$, and [O/H]$=-0.29 \pm 0.07$. In the same work, the age was estimated to be between 15 to 25 Myr \citep{Larsen2008}.

We wanted to see what effect the different ratios of red-to-blue supergiants, $N_{RSG}/N_{BSG}$, would have on the derived chemical abundances. This ratio in NGC 1569-B has been previously discussed in the literature. For example, in \cite{Larsen2008}, the derived ratio was $N_{RSG}/N_{BSG}$ = 2.50$\pm$0.66. However, \cite{Larsen2011} used resolved photometry and identified 69 red SGs (RSGs) and 45 blue SGs (BSGs) giving $N_{RSG}/N_{BSG}$ = 1.53$\pm$0.09. 
The difference between the ratios in the \cite{Larsen2008} and \cite{Larsen2011} studies is mainly due to the way the BSGs were selected, which illustrates that it is challenging to separate BSGs and the bright main sequence (MS) stars into two distinguishable groups. 

Here, we study the chemical composition of YMC NGC 1569-B, using a high-resolution optical IL spectrum and applying the same method and software created by \cite{Larsen2012}.
We derive abundance ratios for a larger set of elements than those previously measured, including several $\alpha$-elements (Mg, Si, Ca, Ti), Fe-peak elements (Sc, Cr, Mn, Ni) and the neutron capture element:\ Ba. 
Section\,\ref{data} presents the observations we used, which are described together with the data reduction pipeline. The main stages of the analysis method are explained in Section\,\ref{methods}, including the investigation of an important uncertainty for this study: the red-to-blue supergiant ratio (Section\,\ref{RBSG}). The results are given in Section\,\ref{results}. We discuss and compare our findings to other YMCs, the MW stars, and LMC stars in Section\,\ref{discussion}. We offer a discussion of the supergiant ratios in Section\,\ref{RBSG_discussion}. In Section\,\ref{conclusions}, we present our main conclusions.

\section{Data}
\label{data}
\subsection{IL spectrum}
NGC 1569-B was observed with the high-resolution echelle spectrometer (HIRES) on the Keck I 10-m telescope on the 1st of November 2018. 
The exposure time was $2\times1800$~s with the C5 decker, a $7\arcsec\times1\farcs148$ slit. This gives a resolving power of $R = \lambda/\Delta \lambda = 37500$. The
right ascension and declination are 04:30:49.01 and +64:50:52.7 (J2000), respectively. 
These data were reduced using the highly automated MAKEE package (written by T. Barlow) that was specifically designed for HIRES data.
The pipeline performs bias subtraction and flat fielding, determines the location of the echelle orders on the images and extracts the spectra. The spectra of Th-Ar calibration lamps mounted in HIRES were used to calibrate the wavelength. 
Afterwards, the reduced two spectra were used to produce the final IL spectrum and an error spectrum. The final spectrum consists of a sum of two observations covering the $\approx$3900-8150 \AA\ wavelength range; whereas the error spectrum is a difference between the two observations, providing an uncertainty for every pixel in the spectrum. The error spectrum was convolved with a boxcar kernel, using the running mean with window of N=100 pixels, this helped to remove the anomalies from the error spectrum, such as telluric absorption (TA), cosmic rays, and so on. The signal-to-noise ratio (S/N) increases from the blue to the red end of the spectrum, with typical values of $\sim$50 at 4000 \AA\ and $\sim$70 at 6000 \AA.

\subsection{HST photometry}
The photometry data were obtained with the Advanced Camera for Surveys (ACS) onboard HST, for the programme GTO-9300 (P.I.: H. Ford).
This provided us with F555W and F814W bands, which can be approximated to V and I. The exposure time was 390 s (3 $\times$ 130) for both filters. These data were previously used in \cite{Larsen2008} and the photometry is taken directly from it. As described in this paper, the photometry was foreground extinction-corrected and is not insignificant in the blue-optical ($A_V$ =1.74 \cite{Israel1988}). The available NGC1569-B photometry was used to find the best-fit isochrone, indicating an age of $\sim$25 Myr. The isochrone was used to perform the photometry test to see the contribution of light from red and blue supergiants (see Section\,\ref{RBSG}).

\section{Methods} 
\label{methods}
\subsection{Models and stellar parameters}

The analysis of high-resolution IL data for star clusters is described in detail in \cite{Larsen2012, Larsen2014, Hernandez2017}.
Here, we give a brief overview of the steps particularly relevant to the present study.

The first step in the analysis is to split the cluster HRD into bins (up to 100) of different stellar evolutionary stages. This can be done using a combination of the isochrone and available photometry or the isochrones alone. Here, only the isochrone was used to create the bins because the photometry of the NGC 1569-B is not sufficiently deep to properly sample the CMD, preventing us from performing an accurate cluster and field separation. Each bin is described by mass, effective temperature ($T_\mathrm{eff}$), logarithm of surface gravity ($\log(g)$), radius of the star, weight of the bin, logarithm of microturbulent velocity ($\log(\nu_t)$), and the atmospheric model used to produce the synthetic spectra. The two different atmospheric models used are: ATLAS9 for stars with $T_\mathrm{eff}>4000$~K \citep{Kurucz1970} and for the cool stars ($T_\mathrm{eff}<4000$~K) MARCS models \citep{Gustafsson2008}. 
The weight of each bin is calculated using the IMF and the mass of stars in the bins. We use the power law for the IMF, $\mathrm{d}N/\mathrm{d}M \propto M^{-\alpha}$, with the exponent $\alpha=2.35$ from \cite{Salpeter1955}.
The microturbulent velocities were allocated depending on the stellar effective temperature (similarly to \cite{Lyubimkov2004, Hernandez2017}): for stars with $T_\mathrm{eff}<6000$ K the microturbulent velocities were $\nu_t$=2.0 km s$^{-1}$; for stars with $6000<T_\mathrm{eff}<22000$ K -- $\nu_t$=4.0 km s$^{-1}$; and stars with $T_\mathrm{eff}>22000$ K had $\nu_t$=8.0 km s$^{-1}$.

Using the Kurucz and MARCS models, the model atmospheres and synthetic spectra are computed for each bin. The codes used for the spectral synthesis are \texttt{SYNTHE} \citep{Kurucz1979} and \texttt{Turbospectrum} \citep{Plez2012}.
Then these spectra are scaled to the stellar luminosity with respect to each bin. Afterwards, these are all combined into a single integrated synthetic spectrum. The abundances used in the modelling are then iteratively adjusted until the best fit to the observed spectra is obtained. 
 
The method we use in this paper is based on the spectra of Arcturus \citep{Wallace2000} and the Sun (2005 version of \cite{Kurucz1984}) to define the spectral windows used for the fitting procedures. A crucial step is to ensure that each window contains lines with a range of equivalent widths, while also including the strong lines (equivalent width > 100 m\AA ). The full and updated list of spectral windows can be found in \cite{Larsen2022}, whereas the list used for this study is presented in \ref{Tab_abun}.
 
Since the method uses full spectral synthesis, blends are automatically accounted for and it is possible to access a wide variety of elements: $\alpha$-abundances (Mg, Si, Ca, Ti), Fe-peak elements (Cr, Mn, Fe, Ni, and Sc), heavy elements (Cu, Zn, Zr, Ba, and Eu), and lighter elements such as Na.

\begin{figure}
\centering
\includegraphics[width=9cm]{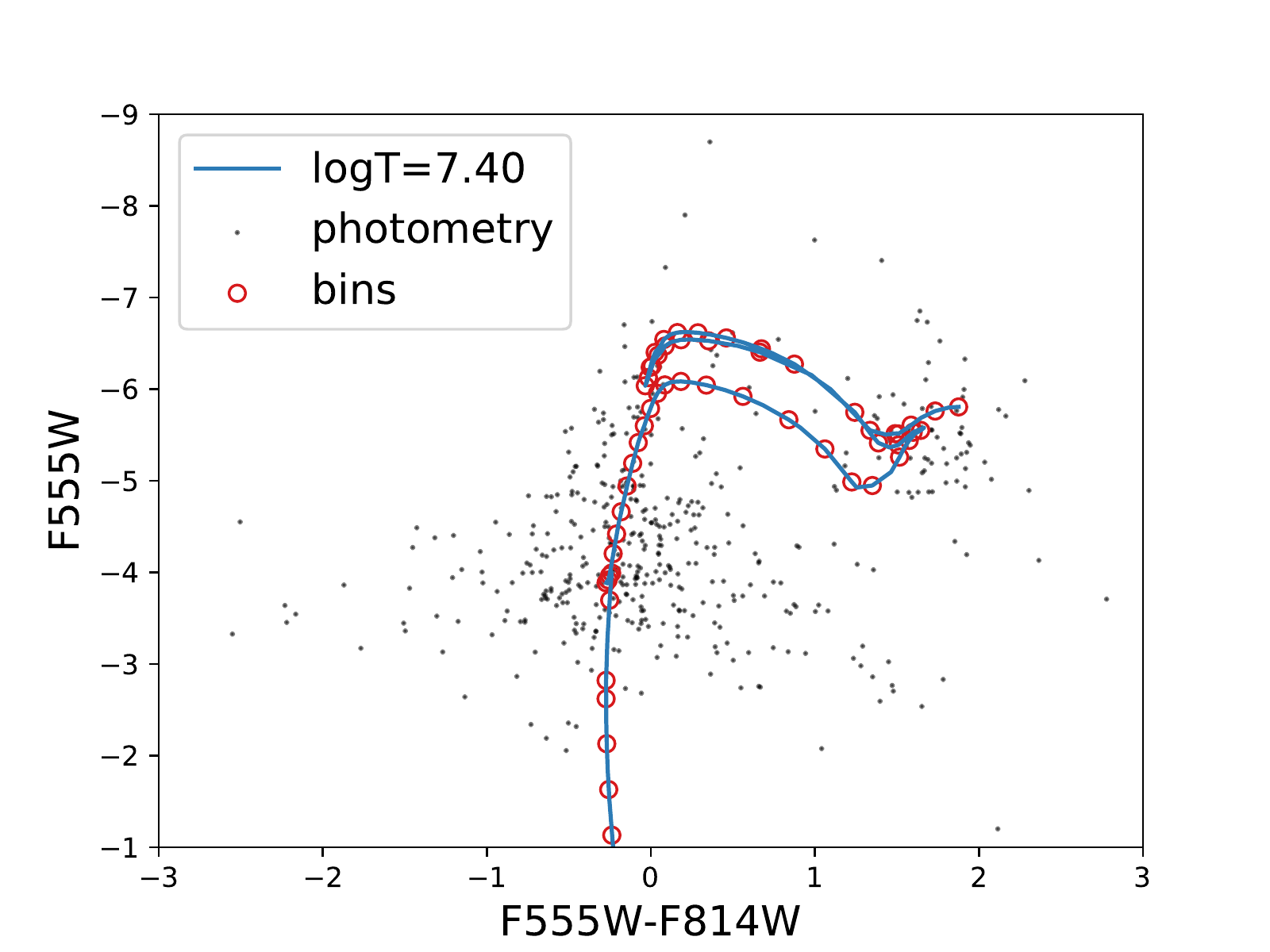}
\caption{CMD of NGC 1569-B using ACS photometry (black dots), with an isochrone (over plotted in blue) of logarithmic age 7.4 (25 Myrs) and Z=0.008 from the Padua catalogue \citep{Girardi2000}. Red empty circles show the theoretical isochrone bins used for the HRD file \citep{Girardi2000}.} 
\label{CMD}
\end{figure}

The blending of the lines is an important detail and a main challenge for the IL spectrum analysis. The \cite{Larsen2012} method was previously applied to high-resolution spectra of GCs, however, young stellar clusters such as NGC~1569-B tend to have a higher degree of blending compared to metal-poor, old GCs due to the higher metallicity (although GCs up to nearly solar metallicity were included in \citealt{Larsen2022}). 
Therefore, to account for this, the analysed windows had to be carefully arranged to ensure that the element lines do not overlap with others produced by different types of elements.
The spectral windows fitted are a combination of the previously used set of wavelength regions \citep{Larsen2022} and a new set of lines for the blue part of the spectrum (Table\,\ref{elem_list}).
    
\subsection{Ratio of red-to-blue supergiants}
\label{RBSG}

One of the major uncertainties that we aim to tackle in this analysis is the impact of the ratio of red-to-blue supergiant stars.
RSG stars are cool stars which are the brightest stars in NIR light for young stellar populations. These stars evolve from the MS as their hydrogen is depleted and heavy elements come closer to the stellar surface. The following step of the stellar evolution is the BSG star phase, which are hot stars and the most luminous in the blue optical part of the spectrum \citep{Pagel2009book}.

\begin{figure*}
\centering
\includegraphics[width=17cm]{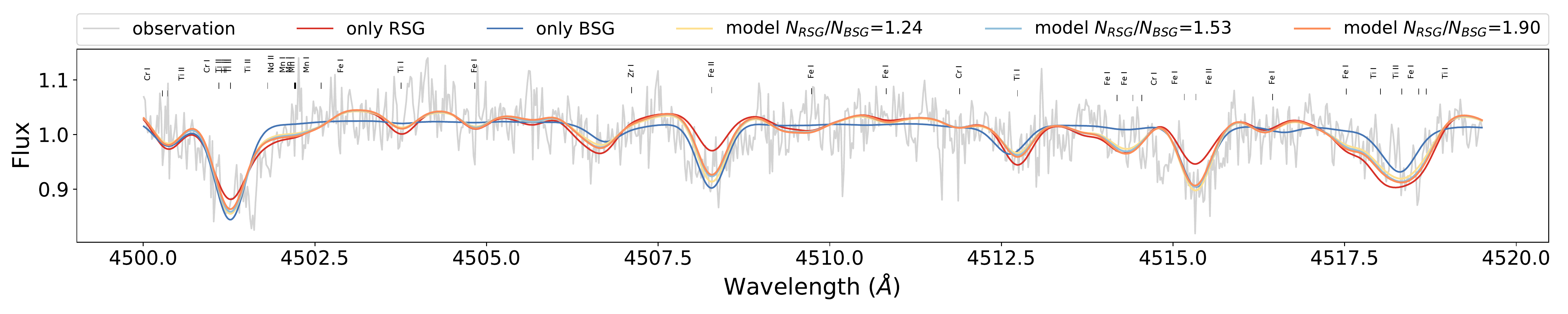}
\includegraphics[width=17cm]{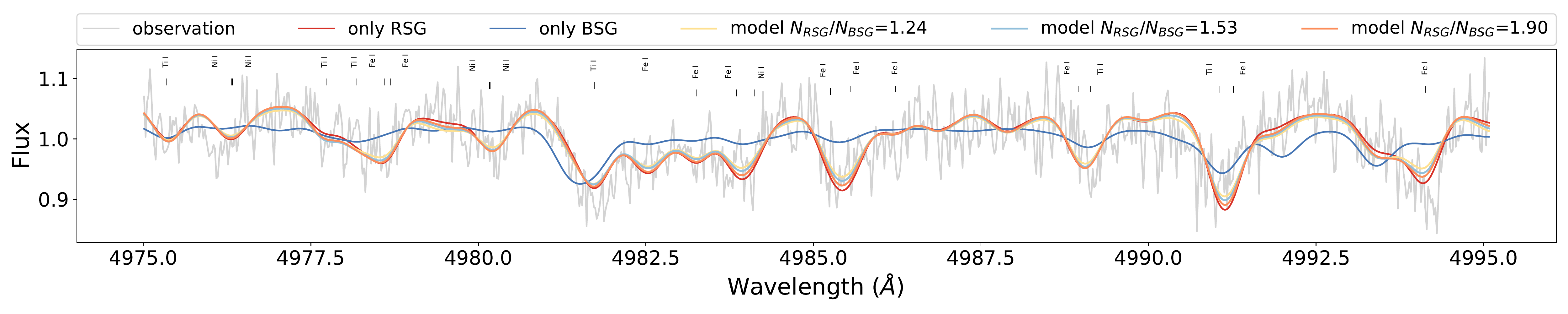}
\includegraphics[width=17cm]{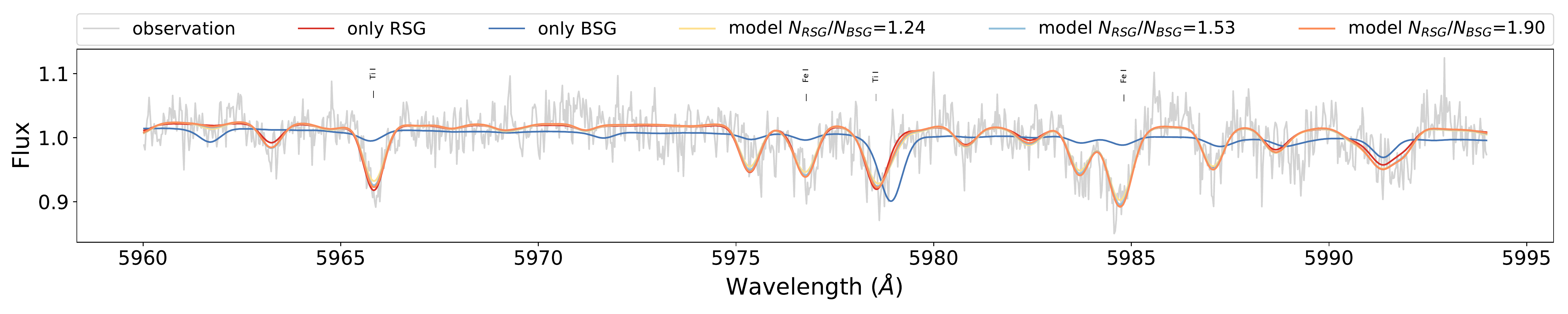}
\caption{Example of the fits for three different Ti windows, grey is the observed spectrum, red is the model with no BSGs, blue is the model with no RSGs, yellow is the model with $N_{RSG}/N_{BSG}$=1.24, light blue is the model with $N_{RSG}/N_{BSG}$=1.53, and orange is the model with $N_{RSG}/N_{BSG}$=1.90.} 
\label{noXSG1}
\end{figure*}

\begin{figure*}
\centering
\includegraphics[width=17cm]{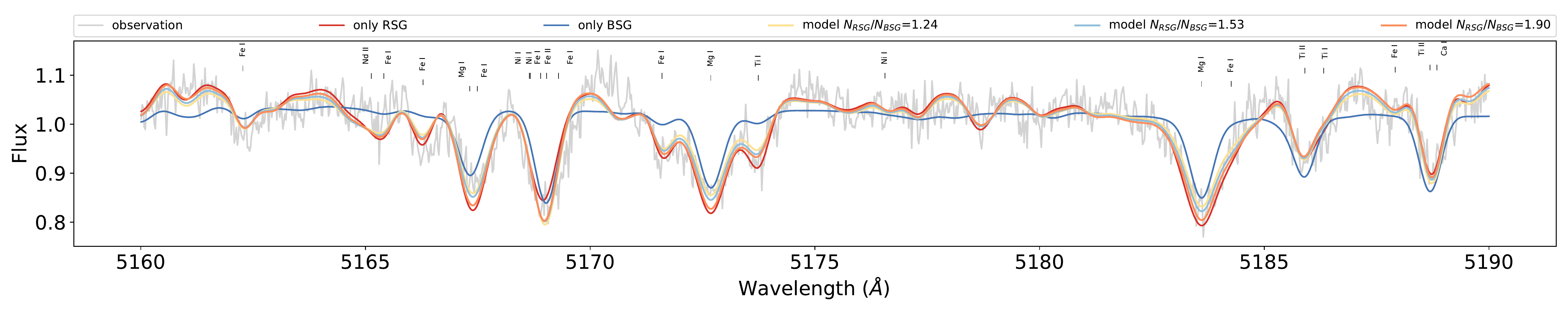}
\caption{Fit for the window with Mg B triplet lines, plotted as in Figure\,\ref{noXSG1}.} 
\label{Mg_spectrum}
\end{figure*}

Given the very different effective temperatures of blue and red SG stars, they affect different parts of the spectrum. This effect is shown in Figures\,\ref{noXSG1} and\,\ref{Mg_spectrum}, which provide a comparison between the best-fit model of a stellar population with only BSG (shown in blue) and with only RSG (shown in red). It can be seen that at shorter wavelengths, the blue model spectrum tends to provide a better match overall to the observations; while at longer wavelengths, the RSGs with their more numerous features start dominating the integrated light.
The wavelength range of the analysed IL spectrum is $\sim 4000-8000$ \AA\ and the main contribution in the blue up to about 5500 \AA\ is from BSGs; while the red part of the spectrum is more dominated by the light from RSGs. 
Since the light from RSGs dominates the spectrum at redder wavelengths and has almost no effect at the shorter wavelengths (see Section\,\ref{Subsec:ectfract_of_lightRBSG}), by increasing $N_{RSG}/N_{BSG}$ the IL spectrum will become redder. As expected, the BSGs have the opposite effect on the spectrum, and decreasing the ratio will mainly affect the bluer part of the spectrum.
These wavelength ranges indicate that the optimal contribution between the blue and the red SG stars has to be achieved. 

\begin{figure}
\centering
\includegraphics[width=9cm]{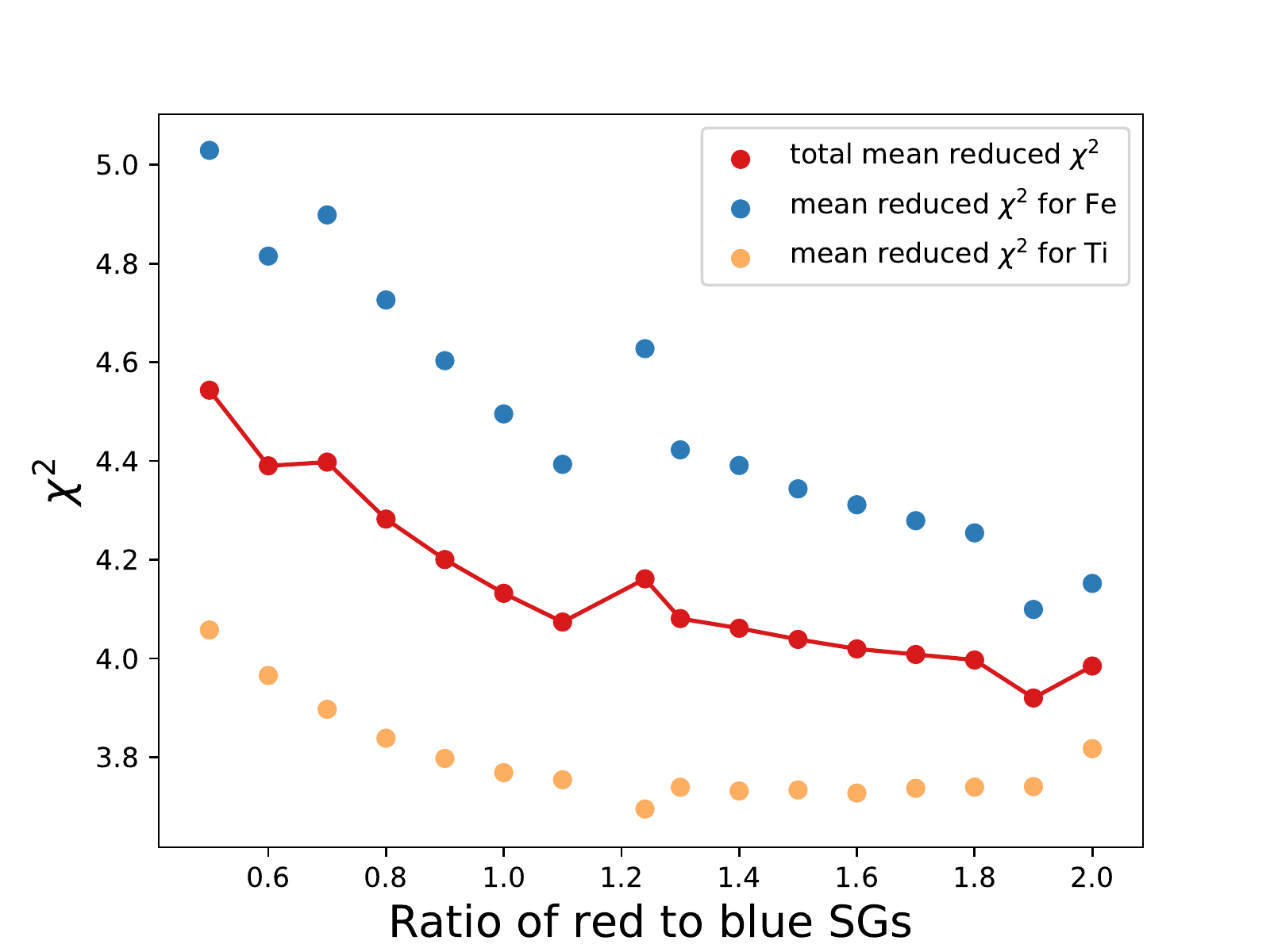}
\caption{Different ratios of $N_{RSG}/N_{BSG}$ versus the reduced $\chi^2$ values derived for the runs with these red-to-blue SG stars ratios. $\chi^2$ values were summed over all Ti and Fe lines that were fitted. The values of Fe $\chi^2$ best model fits are in blue, Ti $\chi^2$ values in orange, and the combined $\chi^2$ values in red.} 
\label{chi_ratio}
\end{figure}

The RSGs and BSGs bins were selected using the surface gravity and effective temperature from an isochrone: $\log(g)\leq2.5$ and $T_\mathrm{eff}\leq5000$ K for RSGs, while $\log(g)\leq2.5$ and $T_\mathrm{eff}\geq5000$ K for BSGs. The number of blue and red SGs were then calculated using the weight of each bin based on the IMF \citep{Salpeter1955}.
The original ratio within the isochrone used for 25 Myrs and Z=0.008 from the 2008 version of the Padua isochrones has $N_{RSG}/N_{BSG} = 1.24$ \citep{Girardi2000, Girardi2008, Marigo2008}.
However, another way to find the optimal red-to-blue SG ratio is to use statistics on the existing spectroscopic data.
It is possible to combine the reduced $\chi^2$ values of multiple elements best fits to calculate the total $\chi^2$ value characterising the runs with different red-to-blue SGs ratios. Figure\,\ref{chi_ratio} shows how $\chi^2$ varies for ratios between 0.5 and 2.0, namely, higher ratios between 2.0 and 4.0 fail to fit multiple Fe lines; hence, those ratios were not considered in the search for the lowest $\chi^2$. A  visible trend is that the higher ratios of RSGs to BSGs are favoured by the analysis. The ratio with the lowest $\chi^2$ value and therefore the most statistically accurate is $N_{RSG}/N_{BSG}$=1.90.

\begin{table}
\caption{List of newly added lines to the analysis.} 
\label{elem_list}

\centering

\begin{tabular}{cll} 
\hline\hline             
Element/Molecule & Wavelength [\AA] & Reference \\
\hline
  Ti & 4293.0-4315.0 & 1 \\
    & 4397.0-4429.0 & 1 \\
    & 4442.0-4475.0 & 1 \\  
    & 4521.0-4540.0 & 1 \\
    & 4570.0-4575.0 & 1 \\
  TiO1 & 5960.0-5994.0 & 2 \\
  
  Si & 4367.0-4377.0 & 1 \\ 
   & 4382.0-4397.0 & 1 \\
   & 4420.5-4430.5 & 1 \\
   & 4596.0-4606.0 & 1 \\
   & 4622.0-4632.0 & 3  \\
   & 4701.0-4711.0 & 1 \\
   & 4716.0-4726.0 & 3 \\
   & 4743.0-4753.0 & 1 \\
   & 4942.0-4952.0 & 1 \\
   
  Cr & 4253.0-4260.0 &  1\\
    & 4270.0-4276.0 & 1\\
    & 4565.0-4570.0 & 1\\
    & 4578.0-4597.0 & 1\\

  \\

\hline
\end{tabular}
\tablebib{(1)~\cite{Kudritzki2014} and references therein, (2)~\cite{Nuria2021}, (3)~\cite{Herrero2020}.}

\end{table}

\section{Results}
\label{results}
\subsection{Chemical abundances}
We derived the abundances for a number of individual elements from the NGC 1569-B IL spectrum, starting with the elements with the highest number of lines.
The first step was to fit for the broadening and overall
scaling of the solar reference abundances, [Z/H], modelling the whole wavelength range for every order. Once derived, the broadening can be fixed for all the subsequent runs of this YMC. The first individual element to fit is Fe since it has the largest number of absorption lines, we subsequently fit for Ti and so on. This procedure allows us to obtain the individual element abundances values for NGC 1569-B (Table\,\ref{Tab_abun}). Table\,\ref{fullabun} lists all the values of the averaged abundances derived for three different ratios of red to blue SGs:  1.24 – the original ratio from the isochrone, 1.53 – the observed ratio from \cite{Larsen2011}, and 1.90 – based on the best $\chi^2$ value from iteratively fitting the spectrum (Figure\,\ref{chi_ratio}).

\begin{figure}
\centering
\includegraphics[width=9cm]{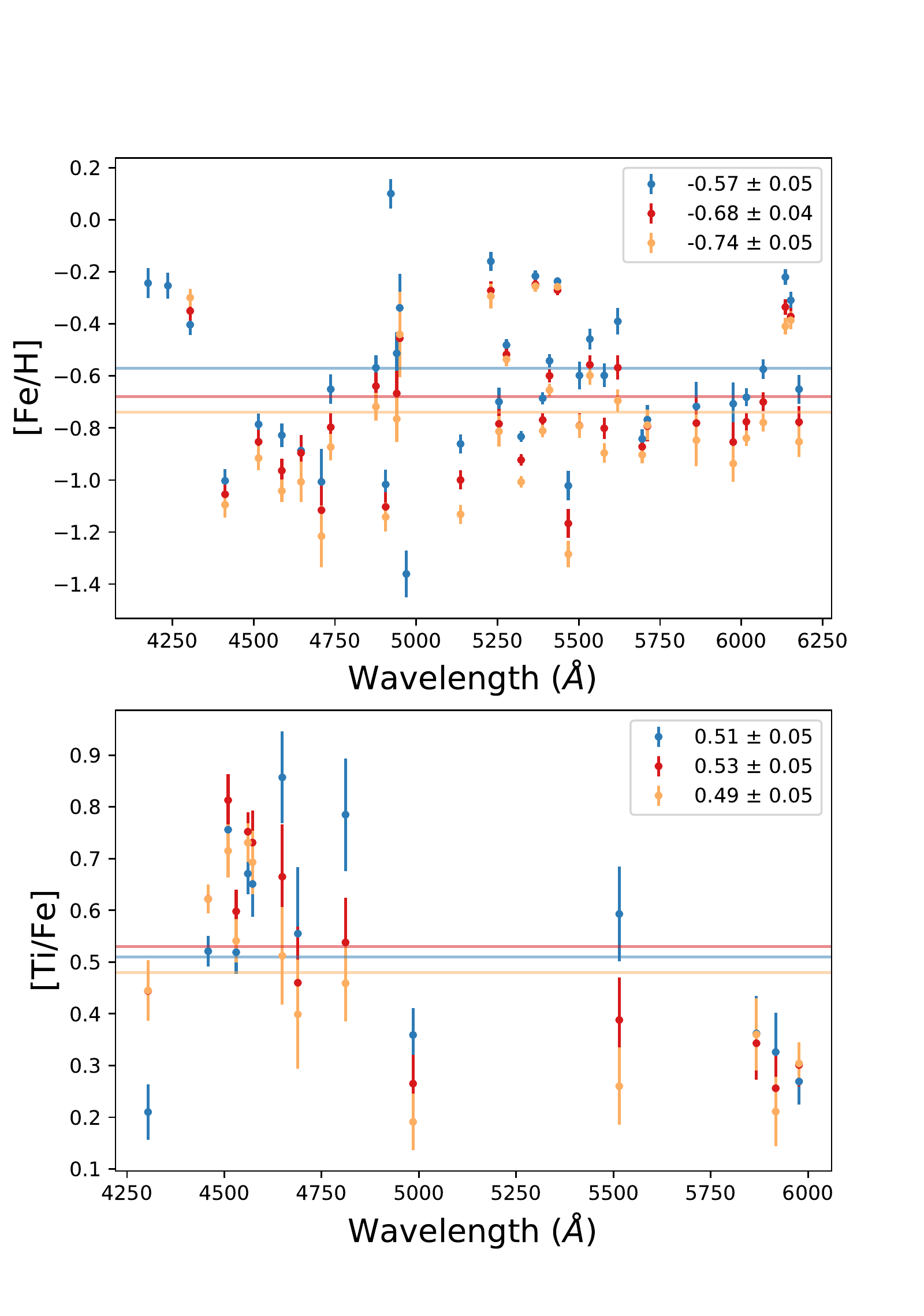}
\caption{[Fe/H] abundances derived for each of the fitted wavelength windows, shown at the top. The blue dots represent $N_{RSG}/N_{BSG}$=1.24, the red points - $N_{RSG}/N_{BSG}$=1.53, and yellow - $N_{RSG}/N_{BSG}$=1.90 runs. The lines of corresponding colours indicate the averaged values of [Fe/H] for there runs.  [Ti/Fe] abundances derived for each of the fitted wavelength regions shown at the bottom. The blue dots represent $N_{RSG}/N_{BSG}$=1.24, the red points - $N_{RSG}/N_{BSG}$=1.53, and yellow - $N_{RSG}/N_{BSG}$=1.90 runs. The lines of corresponding colours indicate the averaged values of [Ti/Fe] for there runs. 
} 
\label{fe_ti}%
\end{figure}

Section\,\ref{Subsec:ectfract_of_lightRBSG} shows that the isochrones reproduce the observed fraction of light from the RSGs quite well. However, for this particular spectral data the $\chi^2$ argument favours the higher ratio. For some elements, there is some disagreement between the blue and the red part of the spectrum (see the bottom plot in Figure\,\ref{fe_ti} or the numerical values for Ca and Cr in the Table\,\ref{Tab_abun}). In general, the scatter in abundances increases towards the blue where the light is increasingly dominated by BSG and MS stars. 
The uncertainties on the final abundance ratios are the weighted standard deviation average;
this way, all the scatter is accounted for (see Section\,\ref{Section:Uncertainties}). The separate abundance values for every fitted wavelength window are listed in the Table\,\ref{Tab_abun}. Windows for which the spectral fitting procedure failed to converge to a value within an allowed margin (e.g. $-1 < \mathrm{[X/Fe]} < +1$) are marked as “-”.
The individual errors tabulated are the formal errors based on the $\chi^2$ analysis.

We derive sub-solar iron abundance ratios of [Fe/H]$_{1.24}=-0.57\pm0.05$, [Fe/H]$_{1.53}=-0.68\pm0.04$, and [Fe/H]$_{1.90}=-0.74\pm0.05$ for 1.24, 1.53, and 1.90 ratios, respectively\footnote{From here onwards the index at the bottom of the element abundance indicates the ratio of red-to-blue SGs used to derive this value.}. 
We have used 38 windows containing lines of \ion{Fe}{i} and \ion{Fe}{ii} including the new wavelength regions to include the lines from the blue part of the spectrum (see Table.\ref{elem_list}).
The $\alpha$-abundances, [<Mg, Si, Ca, Ti>/Fe], for all of the ratios are 'super-solar', with a mean $\alpha$-element enhancement of [$\alpha$/Fe]$_{1.90}=+0.25 \pm$0.11.

The abundances of Fe-peak elements Cr, Mn, Ni, and Sc were determined from a collection of spectral lines. [Mn/Fe] and [Ni/Fe] are sub-solar and slightly super-solar, respectively ($\text{[Mn/Fe]}=-0.22\pm0.12$ and $\text{[Ni/Fe]}=+0.13 \pm 0.11$), while [Cr/Fe] and [Sc/Fe] abundance ratios are found to be super-solar. In particular, Sc appears to be high for all three values of $N_{RSG}/N_{BSG}$: [Sc/Fe]$_{1.24}=+0.95\pm0.16$, [Sc/Fe]$_{1.53}=+0.87\pm0.19$, and [Sc/Fe]$_{1.90}=+0.78\pm0.20$. The value of [Cr/Fe] is derived to be $+0.50\pm0.11$.
Cu was determined using the 5105 {\AA} \ion{Cu}{i} line, while the other frequently used line at 5782 {\AA} coincides with a diffuse interstellar band and was not included.

Four windows with \ion{Ba}{ii} lines, at 4554, 5854, 6142, and 6497 {\AA}, were used to determine [Ba/Fe]. The resulting Ba abundance ratios indicate an extremely super-solar ratio, $\text{[Ba/Fe]}=+1.28\pm0.14$.

\begin{figure*}
\centering
\includegraphics[width=15cm]{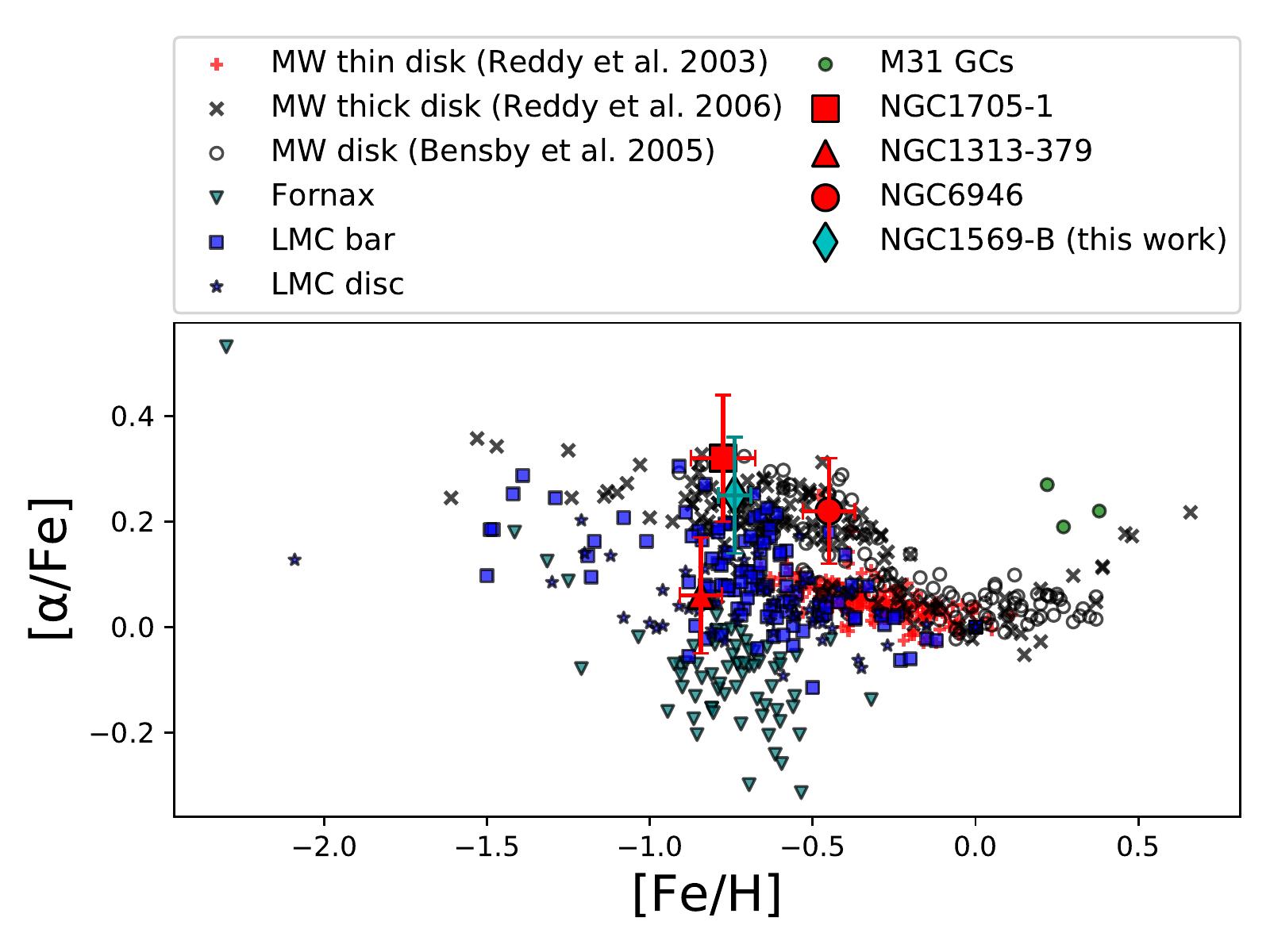}
\caption{[$\alpha$/Fe] plotted against iron abundance, [Fe/H].
Turquoise rhombus refers to this work; 
red triangles show the alpha abundances estimated for NGC 1313-379; 
red squares display the measurements for NGC 1705-1 \citep{Hernandez2017}; 
red circles display the measurements for the SSC in NGC 6946 \citep{Larsen2006} ; 
green circles show the abundance of the M31 GCs from \cite{Colucci2009};
red crosses and black Xs present MW disc abundances from \cite{Reddy2003, Reddy2006}, respectively; 
black open circles are the MW disc abundances from \cite{Bensby2005}; 
teal triangles are the abundances of the individual stars in the centre of the Fornax dwarf spheroidal galaxy \cite{Letarte2010};
blue squares and stars belong to LMC bar and inner disc abundances presented by \cite{VanderSwaelmen2013}).} 
\label{alpha_fe}%
\end{figure*}

\begin{figure*}
\centering
\includegraphics[width=18cm]{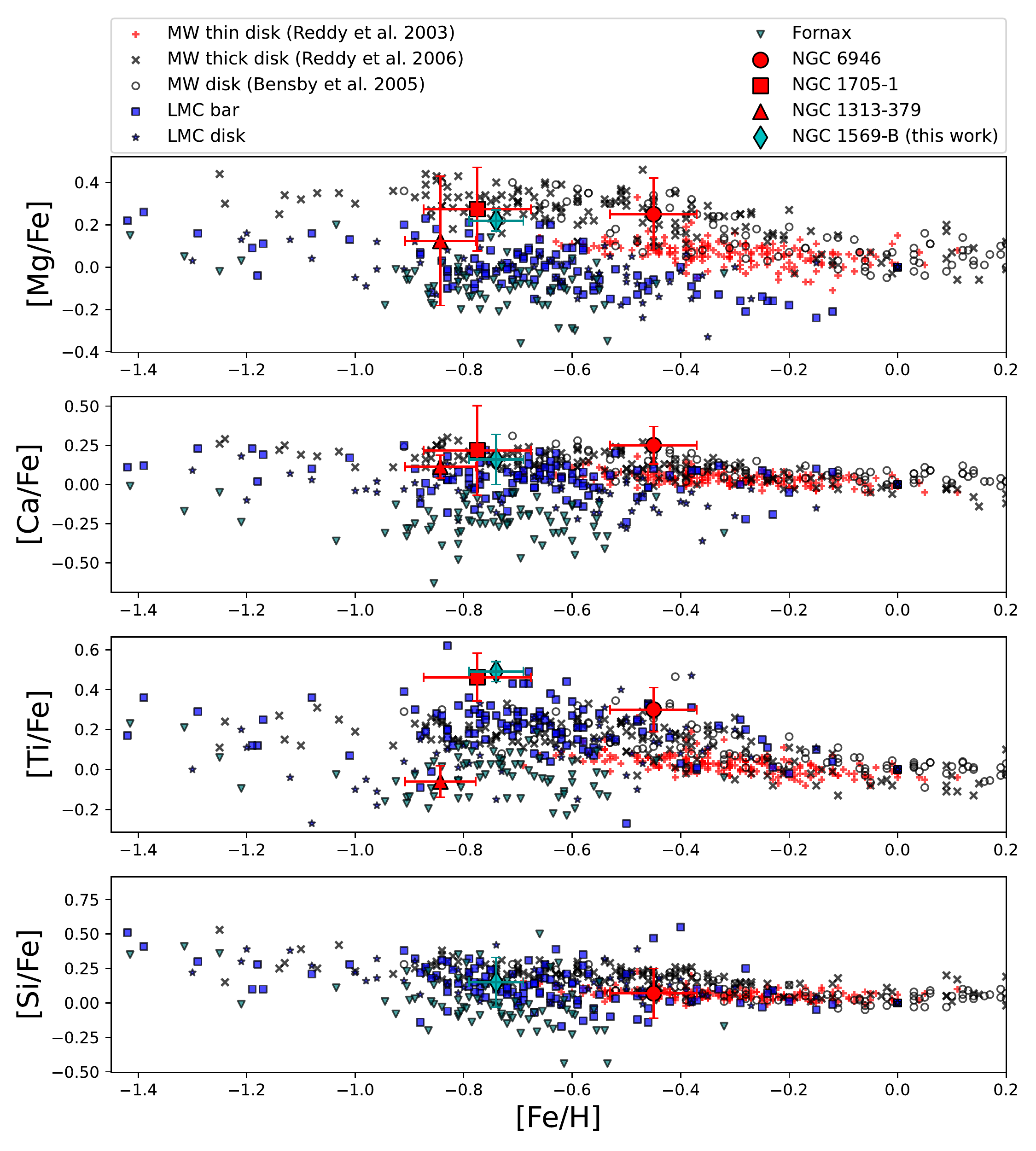}
\caption{Individual $\alpha$-elements plotted against iron abundance, [Fe/H]. Top: [Mg/Fe] vs. [Fe/H]. Second row: [Ca/Fe] vs. [Fe/H]. Third row: [Ti/Fe] vs. [Fe/H]. Bottom: [Si/Fe] vs. [Fe/H]. Symbols as in Figure\,\ref{alpha_fe}.
} 
\label{Fig_alpha_elemVSfe}%
\end{figure*}

\begin{figure*}
\centering
\includegraphics[width=15.2cm]{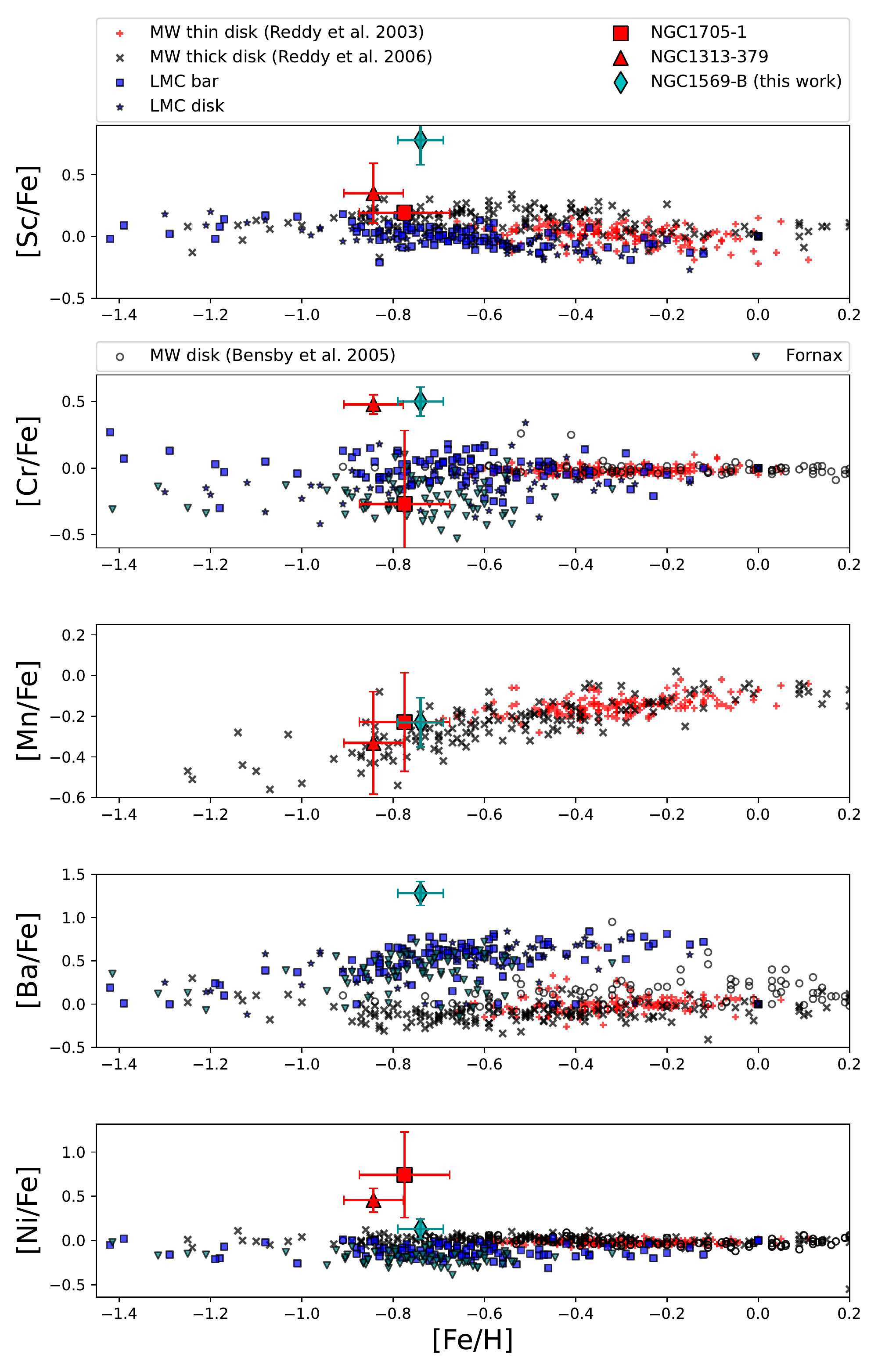}
\caption{Fe-peak and heavy elements plotted against iron abundance, [Fe/H]. Top: [Sc/Fe] vs. [Fe/H]. Second row: [Cr/Fe] vs. [Fe/H]. Third row: [Mn/Fe] vs. [Fe/H]. Fourth row: [Ba/Fe] vs. [Fe/H]. Bottom: [Ni/Fe] vs. [Fe/H]. Symbols as in Figure\,\ref{alpha_fe}.
} 
\label{Fig_other_elemVSfe}%
\end{figure*}

\begin{table*}
\caption{Individual element abundances derived for three different ratio of red-to-blue SGs.}
\label{fullabun}

\centering

\begin{tabular}{ccccccc} 
\hline\hline             
$N_{RSG}/N_{BSG}$ & [Fe/H] & [Mg/Fe] & [Ca/Fe] & [Ti/Fe] & [Si/Fe] & [$\alpha$/Fe] \\
\hline

  1.24 & -0.57 $\pm$ 0.05 & +0.04 $\pm$ 0.10 & +0.42 $\pm$ 0.14 & +0.51 $\pm$ 0.05 & +0.04 $\pm$ 0.15 & +0.25 $\pm$ 0.11
  \\
  1.53 & -0.68 $\pm$ 0.04 & +0.13 $\pm$ 0.08 & +0.32 $\pm$ 0.16 & +0.53 $\pm$ 0.05 & +0.09 $\pm$ 0.18 & +0.27 $\pm$ 0.12
  \\
  1.90 & -0.74 $\pm$ 0.05 & +0.22 $\pm$ 0.05 & +0.16 $\pm$ 0.16 & +0.49 $\pm$ 0.05 & +0.15 $\pm$ 0.18 & +0.25 $\pm$ 0.11
  \\
  \hline
  $N_{RSG}/N_{BSG}$ & [Cr/Fe] & [Mn/Fe] & [Sc/Fe] & [Ba/Fe] & [Ni/Fe] & [Cu/Fe]\\
  \hline
  1.24 & +0.61 $\pm$ 0.12 & +0.03 $\pm$ 0.17 & +0.95 $\pm$ 0.16 & +1.28 $\pm$ 0.21 & +0.40 $\pm$ 0.14 & +0.21 $\pm$ 0.17
  \\
  1.53 & +0.55 $\pm$ 0.11 & -0.17 $\pm$ 0.14 & +0.87 $\pm$ 0.19 & +1.27 $\pm$ 0.19 & +0.23 $\pm$ 0.12 & -0.03 $\pm$ 0.19 
  \\
  1.90 & +0.50 $\pm$ 0.11 & -0.22 $\pm$ 0.12 & +0.78 $\pm$ 0.20 & +1.28 $\pm$ 0.14 & +0.13 $\pm$ 0.11 & -0.17 $\pm$ 0.18 
  \\
  
\hline
\end{tabular}
\tablefoot{The listed errors are weighted standard deviation average ($ S_{\text{X}} $).}
\end{table*}

\subsection{Uncertainties}
\label{Section:Uncertainties}

Each element had a number of spectral windows fitted. Table\,\ref{Tab_abun} provides the full list of wavelength ranges, the derived values, and their uncertainties from iSPy3. Those values were then combined to produce the final value for single elements.
The values and uncertainties listed in the Table\,\ref{fullabun} are the weighted average of the values derived for the spectral windows for every individual element:

\begin{equation}
\langle [\text{X/Fe}] \rangle=\frac{\sum w_i[\text{X/Fe}]_i}{\sum w_i}
,\end{equation}

where the weights are defined as

\begin{equation}
w_i=\frac{1}{\sigma_i^2+\sigma_0^2}
,\end{equation}

and where $\sigma_0 = 0.05$ dex is used to account for non-random uncertainties on the derived individual window abundances. Then the uncertainties were calculated using the following formula:

\begin{equation}
\sigma_{\langle \text{X} \rangle}=\left( \sum w_i \right)^{-1/2}
.\end{equation}

A different approach is to take the standard errors ($S_{\text{X}}$) on the mean of the weighted standard deviations ($SD_{X,w}$): 

\begin{equation}
\text{SD}_{X,w}= \left[ \frac{N}{N-1} \frac{\sum (\text{X/Fe}_i - \langle \text{X/Fe} \rangle )^2 w_i}{\sum w_i} \right]^{1/2}
,\end{equation}
\begin{equation}
S_{\text{X}}=SD_{X,w}/ \sqrt{N}
.\end{equation}

The comparison between the random and systematic uncertainties are listed in Table\,\ref{tab:error table}. In particular, $\sigma_{\langle \text{X} \rangle}$ is always lower than $S_{\text{X}}$, except for the errors of Cu (this element had a single spectral window). This indicates that the errors calculated using the weighted average tend to underestimate the uncertainties, even if the floor value is used ($\sigma_0$). Therefore, weighted standard deviation average errors are used as a more accurate estimate of the true uncertainties.

\begin{table}
\caption{Error table.}
\label{tab:error table}
\centering
\begin{tabular}{cccc}
\hline
\hline
 Abundance ratio & Weighted average & $\sigma_{\langle \text{X} \rangle}$ & $\text{S}_{\text{X}}$
\\ 
\hline
[Fe/H] & -0.74 & 0.01  & 0.05 
\\
{[Mg/Fe]} & +0.22 & 0.03 & 0.05 
\\
{[Ca/Fe]} & +0.16 & 0.03 & 0.16
\\
{[Ti/Fe]} & +0.49 & 0.02 & 0.05
\\
{[Si/Fe]} & +0.15 & 0.04 & 0.18
\\
{[$\alpha$/Fe]} & +0.25 & 0.03 & 0.11 
\\
{[Cr/Fe]} & +0.50 & 0.02 & 0.11
\\
{[Mn/Fe]} & -0.22 & 0.04 & 0.12
\\
{[Sc/Fe]} & +0.78 & 0.04 & 0.20
\\
{[Ba/Fe]} & +1.28 & 0.06 & 0.14 
\\
{[Ni/Fe]} & +0.13 & 0.03 & 0.11
\\
{[Cu/Fe]} & -0.17 & 0.18 & 0.18
\\
\hline
\end{tabular}
\end{table}

\section{Discussion}
\label{discussion}

Observations of field stars and star clusters indicate that NGC~1569 experienced a major recent burst of SF about 25 Myr ago \citep{Greggio1998, Anders2004}. According to the chemical evolution models from \cite{Romano2006}, following such bursty SF the cluster would be enriched with $\alpha$ elements, which is consistent with the enhanced $\alpha$-element abundances that we measure ([$\alpha \text{/Fe]}=+0.25 \pm 0.11$). Currently, we have no information about abundance variations between the field star populations in NGC 1569 and its YMCs, NGC 1569-A, and NGC 1569-B. Therefore, it would be interesting to search for signatures of a bursty SFH also in other clusters of this dwarf galaxy.

Analyses of the blue part of the spectrum are not a common method for studying young stellar populations such as NGC 1569-B (usually studied using the IR band), due to challenges inherent in modelling BSG stars. This aspect can be clearly seen in Figure\,\ref{fe_ti}, where the scatter in individual elements in blue region is significantly larger than that seen at redder wavelengths.
However, the blue region of the IL spectrum not only provides more lines for abundance analysis, but it also helps with determining and constraining the blue-to-red SGs ratios.
Thus, while it is not perhaps the most efficient wavelength range for abundance analysis, it still gives us additional constraints on the properties of the BSG and MS stars and remains to be fully explored.

\subsection{Iron abundance}
The iron abundance derived previously in the literature for NGC 1569-B, namely, $\text{[Fe/H]}=-0.63\pm0.08$, is in agreement with the iron abundances derived with $N_{RSG}/N_{BSG}$=1.24 and $N_{RSG}/N_{BSG}$=1.53 \citep{Larsen2008}.
The values of [Fe/H] derived decrease as the ratio or red-to-blue SGs increases. This behaviour is expected, as the more the RSGs the stronger the stellar lines 
(see Section\,\ref{RBSG_discussion}). Even though the isochrone reproduces the photometry light (see Section\,\ref{Subsec:ectfract_of_lightRBSG}), the final value for the iron abundance is the one obtained with the ratio 1.90 as it is statistically favoured for these spectroscopic data, $\text{[Fe/H]}=-0.74\pm$0.05.

\subsection{$\alpha$ element abundances}
Generally, [$\alpha$/Fe] is a commonly used tracer for a system's timescale of star-formation. In addition to the dependency on SF timescale, [$\alpha$/Fe] is also sensitive to the ratio of former SNII to SNIa and, therefore, to the ratio of massive stars to intermediate mass binaries. As soon as the SNIa start to contribute, [$\alpha$/Fe] decreases. This happens due to the longer time scale of SNIa compared to SNII, on a plot of [$\alpha$/Fe] vs.\ [Fe/H] this feature is referred to as the 'knee'. 
The ratio decreases because the main products from SNII are $\alpha$-elements, in contrast to SNIa, which mainly produce Fe and only a few $\alpha$-elements \citep{Tolstoy2009}.

Our observations show that NGC 1569-B is $\alpha$-enhanced with a super-solar abundance of [$\alpha\text{/Fe]}=+0.25 \pm $0.11, which resembles the $\alpha$-abundances of NGC 1705-1 and the SSC in NGC 6946 ([$\alpha\text{Fe]}=+0.32\pm$0.12 and [$\alpha\text{/Fe]}=+0.22\pm$0.11) $\alpha$ abundances. 
Figure\,\ref{alpha_fe} shows the $\alpha$-abundance versus the iron abundance.
These abundance ratios indicate a rapid and bursty SF, where chemical enrichment is dominated by SNII, which have a massive progenitor star and a short life \citep{McWilliam1997}.

Individual $\alpha$-elements are consistent with other YMCs previously studied, such as NGC 1705-1 whose abundance ratios [Mg/Fe], [Ca/Fe], and [Ti/Fe] resemble those in NGC 1569-B very closely (see Figure\,\ref{Fig_alpha_elemVSfe}), as expected in the bursty models from \cite{Romano2006}. For [Si/Fe] Figure\,\ref{Fig_alpha_elemVSfe} suggests the similarity with Fornax, LMC, and MW thick disk, however, these stars are older and currently there are no YMCs with a similar [Fe/H]. Therefore, it is more difficult to conclude whether this super-solar value of [Si/Fe] is expected, nevertheless this $\alpha$ element is close to the total $\alpha$-abundance.

In addition to atomic \ion{Ti}{i} and \ion{Ti}{ii} lines, we also fit the TiO1 band, which is sensitive to $T_\mathrm{eff}$ and the IMF. 
The derived Ti abundances  using only the TiO1 molecular band for three different ratios are: [Ti/Fe]$_{1.24}=+0.31\pm$0.04, [Ti/Fe]$_{1.53}=+0.34\pm$0.04, and [Ti/Fe]$_{1.90}=+0.33\pm$0.04, which behaves similarly to the overall $\alpha$-abundance ratio. 
This behaviour of the TiO band compared to the total Ti abundance ratio is expected because as we go redder the RSGs will increasingly dominate, making the analysis less sensitive to the ratio of RSG to BSG stars. TiO1 is slightly more sensitive than TiO2 to [Ti/Fe] \citep{La_Barbera2013}.
This method only uses the IMF from \cite{Salpeter1955}, hence, the above stated values might change if a different IMF is used. However, given that we haven't explored other IMF choices, it is currently unclear how strong the constraint on the IMF from the TiO bands is for NGC~1569-B.

\subsection{Fe-peak element abundances} 
The Fe-peak element abundances are shown in (Figure\,\ref{Fig_other_elemVSfe}). These elements are the heaviest elements produced through thermonuclear reactions in SNIa and massive stars \citep{Kobayashi2020, Minelli2021}.
The abundances derived for the Cr and Mn are close to the values of the extragalactic YMCs. The [Mn/Fe] value of NGC~1569-B and that of NGC 1705-1 are in good agreement, while the [Cr/Fe] abundance more closely resembles that of NGC 1313-379. The value of Ni for NGC~1569-B lies closer to the MW and LMC environments, while the derived values of [Ni/Fe] in NGC 1705-1 and NGC 1313-379 show enrichment in Ni.
The high values of Ni in these YMCs were discussed in \cite{Hernandez2017}, offering two possible explanations:\ either there is a different contribution from massive stars and SNIa in these clusters versus the contribution in the MW and LMC or due to the inclusion of a 7700-7800 \AA\ bin, which increased the final value. In particular, if the bin was excluded from the calculation, the [Ni/Fe] values for NGC 1705-1 and NGC 1313-379 would become +0.25 and +0.31, respectively -- instead of +0.74$\pm$0.49 and +0.46$\pm$0.14 (as shown at the bottom plot of Figure\,\ref{Fig_other_elemVSfe}).

The main outlier is the abundance ratio of Sc (produced in both SNIa and SNII), which seems extremely enhanced. 
We considered that this might be caused by an issue in measurement. However, despite changing the fitting combinations of the spline function used for continuum matching of the observed and model spectra (e.g. the power and the number of knots), [Sc/Fe] remained high at all times. 
We see a large range of Sc values across the various spectral window regions which might indicate that our Sc measurements are quite uncertain. For example, if we would exclude the spectral windows at 5522.2-5531.0 \AA\ and 5638.0- 5690.0 \AA,\ the value of [Sc/Fe] would become 0.17$\pm$0.08, which places it very close to the stars of NGC 1705-1 with a Sc abundance ratio of 0.192$\pm$0.052. The analysis of NGC1705-1 had only a single window fitted between 6222.0-6244.0 \AA,\ therefore, there is no overlap in the two analyses for Sc.

Neither of the previously studied SSCs and YMCs have shown such high abundance ratios for this element (see the top plot in Figure\,\ref{Fig_other_elemVSfe}). It is true that Sc is often a tracer of $\alpha$-abundance and while we do find that [$\alpha$/Fe] is enhanced, this does not fully explain the high value of [Sc/Fe].
If the derived enhancement of Sc is real, this would imply that the contribution from SNIa and SNII in NGC~1569-B differs from the contributions in the MW, LMC, and other studied galaxies.

\subsection{Heavy elements abundances}
Heavy elements are those with atomic number Z > 30 (e.g. Ba and Eu) and they are neutron-capture elements. 
In this study, we derived a very high abundance ratio of Ba (the fourth plot in Figure\,\ref{Fig_other_elemVSfe}). Ba is an s-process element, which is produced mainly in low-mass asymptotic giant branch (AGB) stars (1-3 \(M_\odot\)) \citep{Busso1999}. However, it is highly unlikely that AGB stars would be present in a young population of 15-25 Myrs. 
Unlike the case of Sc, all of the fitted spectral windows yield consistently high Ba values, which suggests that it is a more reliable measurement.
One possible explanation for such a high value of [Ba/Fe] could be pollution of the ISM in NGC~1569 coming from AGB stars belonging to earlier bursts of SF.

The Ba abundance is often different from the MW disk stars. For example, in \cite{VanderSwaelmen2013} [BaII/FeI] increases along with an increase of metallicity in the LMC bar, which indicates that LMC chemical enrichment was slower than the one of the MW. Other high values of Ba were obtained for Fornax, which is dominated by slow neutron capture (s-process) at iron-abundance of $\approx-1$; this shows the strong role of the (metal-poor) AGB in its evolution \citep{Letarte2007}. 
A minor contribution of Ba comes from massive stars \citep{Busso1999}.
Conceivably, at low Z, the Ba peak nuclei do not receive the main contribution from the AGB stars, but from the short-lived massive stars instead. These stars produce heavy r-process nuclei that dominate the contribution \citep{Truran1981, Cowan1996, Busso1999}. 

\subsection{Red-to-blue supergiants}

\label{RBSG_discussion}

\subsubsection{Constraints on trends with metallicity from field stars in different environments}
The red-to-blue SGs ratio greatly influences the IL spectra of young stellar populations. However, it is extremely sensitive to details of the stellar evolution models such as the mass loss, convection, and mixing processes, and this has posed a problem in stellar astrophysics for some time \citep{Langer1995}. This ratio was first shown to vary significantly among different galaxies by \citet{Bergh1968}.
\cite{Eggenberger2002} spectroscopically confirmed the increase of this ratio when the metallicity decreases, additionally providing a normalised relation. A study of this ratio in stellar clusters (in the MW, the LMC and the SMC) was performed by \cite{Meylan1982}, who reach a similar conclusion, namely, that the ratio of red-to-blue decreases steeply with increasing metallicity. These authors found a difference in the ratio between the MW and the SMC of approximately an order of magnitude. These results were revisited and corrected for incompleteness by \cite{Humphreys1984} who found that the centre of the MW has a red-to-blue SGs ratio that is ten times lower than of the SMC.
In summary, the red-to-blue SGs ratio is a decreasing function of metallicity, as confirmed by studies in the MW, the Magellanic Clouds, and the Triangulum galaxy, M33 \citep{Robertson1973, Robertson1974, Hagen1974, Ivanov1998}. 

A study of NGC 1569-B using the CMD has shown a significant difference between the observed and the simulated (with single stellar populations (SSP) and multiple-bursts) red-to-blue SG ratios, the latter being three times smaller for NGC 1569-B \citep{Larsen2011}. The number of observationally detected BSGs is affected by the magnitude cuts and the lack of a clear Blue Hertzsprung Gap (BHG) (see below). This makes it challenging to clearly distinguish between BSGs and stars on the upper part of the main sequence. 
\cite{Eldridge2008} compared the observations with their models and found that the best results are achieved by including binary stars; otherwise, either there are too many RSGs, too few BSG, or a combination of the two issues. However, this alone is not enough to explain the observed ratios, and the challenge of the effect of enhanced mass loss and rotation still stands.

In the recent past, the study of red and blue SGs became one of the fundamental ways to explore the extragalactic young stellar populations. This tool allows one to obtain accurate abundance measurements outside of the Local Group \citep{Bresolin2006, Bergemann2013, Gazak2014}. The advantage of working with the RSGs is that the IL spectrum of a stellar cluster can be modelled with a single red SG star, simply because the NIR continuum of young stellar populations is almost entirely dominated by RSG stars \citep{Davies2013}. This approach to modelling the IL as a single RSG was used in \cite{Larsen2008} to study NGC 1569-B in the NIR.

\subsubsection{Fraction of light versus wavelength (NGC 1569-B photometry versus isochrone)}
\label{Subsec:ectfract_of_lightRBSG}
The reason why the red-to-blue SG ratio is crucial for NGC 1569-B is that it is too young to have AGB or RGB stars. Therefore, most of the light of its spectrum comes from the RSGs and BSGs. 
Moreover, the impact of BSGs becomes more important at optical wavelengths, compared to previous studies in the IR, and must be properly taken into account.
A key difficulty in implementing red and blue SGs in this analysis is the disagreement between the stellar evolution tracks and the observations. Evolutionary tracks tend to underestimate the extent of blue loops (He burning stage). The isochrones suggest a gap in the CMD between the H and He burning stellar stages -- the BHG \citep{Mengel1979}. However, this gap is not always clear in the CMD, which results in larger uncertainties of the observed $N_{RSG}/N_{BSG}$, rather than when the uncertainties are predicted by Poissonian errors. 

In our spectral analysis, the ratio of red-to-blue SGs can be varied within the input file containing the stellar population of the cluster, HRD file. We altered the ratio while keeping the total sum of the SG stars constant.
Keeping the total number of SGs constant while varying the $N_\mathrm{RSG}/N_\mathrm{BSG}$ ratio is motivated by the fact that the total amount of time a star spends as a SG (blue or red) is determined mainly by the total luminosity and amount of fuel available. 
Since the (bolometric) luminosity does not vary significantly between the BSG and RSG phases, we then expect that the total number of SGs is roughly independent of the ratio.

To test how well the isochrone reproduces the observed number of RSGs, we calculated the total integrated magnitudes in F555W (V-band) and F814W (I-band) for the observed CMD of NGC~1569-B and for the isochrone used. For NGC~1569-B, this was measured on the HST image within the radial range 15 < r < 50 pixels, where photometry for individual RSG stars was also available. 

The total integrated magnitudes were compared with those of RSGs only, measured in the magnitude range of $21.0 < M_V < 23.6$ and for colours $2.0 < M_V-M_I < 3.0$.
The measured contribution of RSGs was shown to be in full agreement with that predicted by the isochrone in the V-band (22\% and 21\%); whereas for the I-band, it varied slightly: 48\% for the isochrone and 53\% for the photometry. Therefore, for NGC 1569-B, the isochrones reproduce the observed contributions of RSGs to the V- and I-band luminosities quite accurately. However, the fact that stars other than RSGs contribute significantly to the light shows that full SSP modelling is indeed necessary.

\subsubsection{Abundance trends for varying red-to-blue supergiant ratios}

The abundance trends seen with the increase of the $N_{RSG}/N_{BSG}$ ratio act in an expected manner. Particularly there is an apparent decrease in metallicity as $N_{RSG}/N_{BSG}$ increases. This occurs because in general, the spectral lines become stronger as a greater number of RSGs dominate the spectrum. To compensate, the abundances have to decrease to match the observed spectrum. 
However, $\alpha$-abundance ratios relative to Fe stays approximately constant since the abundances of two of the $\alpha$ elements, [Mg/Fe] and [Si/Fe], increase as the ratio increases, while two other element abundances, [Ca/Fe] and [Ti/Fe], decrease. The rest of the element abundance ratios relative to Fe measured tend to decrease
with the increase of the $N_{RSG}/N_{BSG}$.

The ratio of red-to-blue SGs within the isochrone uses an SSP which does not always describe the population of this YMC accurately. For our case, the isochrone ratio was not the optimal one since it lead to significantly poorer solutions in the spectral fit although it reproduced the observed fraction of light from RSGs quite well. We also explored $N_{RSG}/N_{BSG}$ derived from the photometry in \cite{Larsen2011}. However, this photometry was not of ideal quality, hence, the task of resolving and sifting the members of the cluster was challenging. The final ratio used was derived from the minimum $\chi^2$ method using the spectroscopic data available. We decided to use the set of abundance ratios derived with this approach (with $N_{RSG}/N_{BSG}$=1.90), because it was determined statistically for the data used in this study. 

The availability of the photometry can help to constrain some of the stellar types, for instance, BSGs or RSGs. However, not all targets have observed photometry or it is simply not possible to obtain the photometric data. For these cases, the solution with a total $\chi^2$ of varying the ratio of the most luminous stellar types (such as BSGs and RSGs for the YMCs) might help in determining the favoured and optimal solution. Additionally, a comparison among the blue and the red sides of spectrum might provide a useful route to determine the ratio in the integrated spectrum. 

\subsection{Multiple populations}
The detailed chemical composition of a YMC can indicate the presence of MP, for example, Al variations or a strong [Na/Fe] enhancement \citep{Lardo2017, Bastian2020}. While neither of these elements were measured for NGC 1569-B, another sign for MP could be a lower [Mg/Fe] abundance when compared to [Ca/Fe] and [Ti/Fe] 
\citep{Shetrone1996, Kraft1997, Gratton2004, Colucci2009, Larsen2014}. For this YMC, the derived [Mg/Fe] ratio, +0.22$\pm$0.05 is not significantly lower than [Ca/Fe] and [Ti/Fe], +0.16$\pm$0.16 and +0.49$\pm$0.05, respectively. 
Therefore, we do not find evidence of MPs in NGC~1569-B.

\section{Conclusions}
\label{conclusions}

We performed a high-resolution IL spectrum analysis of the YMC NGC 1569-B. This is the first detailed abundance study of this YMC using the full optical wavelength range. The CMD of this cluster is well fit with an isochrone of 25 Myr. From our analysis of the IL spectrum, we find that NGC~1569-B is slightly metal-poor at $\text{[Fe/H]}=-0.77 \pm 0.05$. The cluster displays $\alpha$ and the majority of the Fe-peak (Cr, Mn and Ni) element abundance ratios typical for MW thick disc or LMC bar stars \citep{Reddy2006, VanderSwaelmen2013}. 
We find that the cluster contains super-solar [Ba/Fe] abundances and it is unclear whether the contribution from massive stars can account for the strikingly high derived value of [Ba/Fe]. 
Another unexpectedly high abundance ratio is [Sc/Fe], one of the Fe-peak elements, usually produced in the SNII and SNIa. Although [$\alpha$/Fe] is enhanced, and Sc is a tracer of $\alpha$-abundance, this does not fully explain the derived abundances. A closer look at the fitting procedure also did not resolve the issue. Thus, we are presently unable to explain the extremely high value of [Sc/Fe].

Overall, this YMC resembles the element abundance ratios of another YMC that was the subject of previous studies, namely, NGC 1705-1, which is in agreement with the similar bursty star formations in both of the galaxies modelled by \cite{Romano2006}.
We note the importance of extending the number of studied YMCs, as, currently, the chemical composition of only four extragalactic YMC have been studied in detail. NGC~1569-B is particularly interesting due to its mass since clusters of  this nature are rare in the MW and even in the Local Group. Such massive clusters may provide insights into the multiple populations seen in GCs.
The biggest challenge for the analysis of YMCs in the blue spectral region is the proper treatment of the relative contributions of RSGs and BSGs. These stars have a major impact on the IL spectra of young clusters, hence, they must be  modelled with care. While it remains challenging, the study of the optical IL of YMCs offers potentially important insights into the evolution of this stage of clusters and their host galaxies.

\begin{acknowledgements}
     The data presented herein were obtained at the W. M. Keck Observatory, which is operated as a scientific partnership among the California Institute of Technology, the University of California and the National Aeronautics and Space Administration. The Observatory was made possible by the generous financial support of the W. M. Keck Foundation.
     
     We thank the anonymous referee for a careful and critical reading of the manuscript.
     
     M.A.B. acknowledges financial support from the grant PID2019-107427GB-C32 from the Spanish Ministry of Science, Innovation and Universities (MCIU) and from  the  Severo  Ochoa  Excellence  scheme (SEV-2015-0548). This work was backed in part through the IAC project TRACES which is  supported through the state budget and the regional budget of the Consejería de Economía, Industria, Comercio y Conocimiento of the Canary Islands Autonomous Community.
     
\end{acknowledgements}

\bibliographystyle{aa}
\bibliography{43415corr}

\begin{thebibliography}{78}
\expandafter\ifx\csname natexlab\endcsname\relax\def\natexlab#1{#1}\fi

\bibitem[{{Aloisi} {et~al.}(2001){Aloisi}, {Clampin}, {Diolaiti}, {Greggio},
  {Leitherer}, {Nota}, {Origlia}, {Parmeggiani}, \& {Tosi}}]{Aloisi2001}
{Aloisi}, A., {Clampin}, M., {Diolaiti}, E., {et~al.} 2001, \aj, 121, 1425

\bibitem[{{Anders} {et~al.}(2004){Anders}, {de Grijs}, {Fritze-v. Alvensleben},
  \& {Bissantz}}]{Anders2004}
{Anders}, P., {de Grijs}, R., {Fritze-v. Alvensleben}, U., \& {Bissantz}, N.
  2004, \mnras, 347, 17

\bibitem[{{Bastian} \& {Lardo}(2018)}]{Bastian2018}
{Bastian}, N. \& {Lardo}, C. 2018, \araa, 56, 83

\bibitem[{{Bastian} {et~al.}(2020){Bastian}, {Lardo}, {Usher}, {Kamann},
  {Larsen}, {Cabrera-Ziri}, {Chantereau}, {Martocchia}, {Salaris}, {Asa'd}, \&
  {Hilker}}]{Bastian2020}
{Bastian}, N., {Lardo}, C., {Usher}, C., {et~al.} 2020, \mnras, 494, 332

\bibitem[{{Bensby} {et~al.}(2005){Bensby}, {Feltzing}, {Lundstr{\"o}m}, \&
  {Ilyin}}]{Bensby2005}
{Bensby}, T., {Feltzing}, S., {Lundstr{\"o}m}, I., \& {Ilyin}, I. 2005, \aap,
  433, 185

\bibitem[{{Bergemann} {et~al.}(2013){Bergemann}, {Kudritzki}, {W{\"u}rl},
  {Plez}, {Davies}, \& {Gazak}}]{Bergemann2013}
{Bergemann}, M., {Kudritzki}, R.-P., {W{\"u}rl}, M., {et~al.} 2013, \apj, 764,
  115

\bibitem[{{Bresolin} {et~al.}(2006){Bresolin}, {Pietrzy{\'n}ski}, {Urbaneja},
  {Gieren}, {Kudritzki}, \& {Venn}}]{Bresolin2006}
{Bresolin}, F., {Pietrzy{\'n}ski}, G., {Urbaneja}, M.~A., {et~al.} 2006, \apj,
  648, 1007

\bibitem[{{Busso} {et~al.}(1999){Busso}, {Gallino}, \&
  {Wasserburg}}]{Busso1999}
{Busso}, M., {Gallino}, R., \& {Wasserburg}, G.~J. 1999, \araa, 37, 239

\bibitem[{{Cabrera-Ziri} {et~al.}(2016){Cabrera-Ziri}, {Lardo}, {Davies},
  {Bastian}, {Beccari}, {Larsen}, \& {Hernandez}}]{Cabrera-Ziri2016}
{Cabrera-Ziri}, I., {Lardo}, C., {Davies}, B., {et~al.} 2016, \mnras, 460, 1869

\bibitem[{{Cabrera-Ziri} {et~al.}(2020){Cabrera-Ziri}, {Speagle},
  {Dalessandro}, {Usher}, {Bastian}, {Salaris}, {Martocchia},
  {Kozhurina-Platais}, {Niederhofer}, {Lardo}, {Larsen}, \&
  {Saracino}}]{Cabrera-Ziri2020}
{Cabrera-Ziri}, I., {Speagle}, J.~S., {Dalessandro}, E., {et~al.} 2020, \mnras,
  495, 375

\bibitem[{{Colucci} {et~al.}(2009){Colucci}, {Bernstein}, {Cameron},
  {McWilliam}, \& {Cohen}}]{Colucci2009}
{Colucci}, J.~E., {Bernstein}, R.~A., {Cameron}, S., {McWilliam}, A., \&
  {Cohen}, J.~G. 2009, \apj, 704, 385

\bibitem[{{Cowan} {et~al.}(1996){Cowan}, {Sneden}, {Truran}, \&
  {Burris}}]{Cowan1996}
{Cowan}, J.~J., {Sneden}, C., {Truran}, J.~W., \& {Burris}, D.~L. 1996, \apjl,
  460, L115

\bibitem[{{Davies} {et~al.}(2010){Davies}, {Kudritzki}, \&
  {Figer}}]{Davies2010}
{Davies}, B., {Kudritzki}, R.-P., \& {Figer}, D.~F. 2010, \mnras, 407, 1203

\bibitem[{{Davies} {et~al.}(2013){Davies}, {Kudritzki}, {Plez}, {Trager},
  {Lan{\c{c}}on}, {Gazak}, {Bergemann}, {Evans}, \& {Chiavassa}}]{Davies2013}
{Davies}, B., {Kudritzki}, R.-P., {Plez}, B., {et~al.} 2013, \apj, 767, 3

\bibitem[{{Eggenberger} {et~al.}(2002){Eggenberger}, {Meynet}, \&
  {Maeder}}]{Eggenberger2002}
{Eggenberger}, P., {Meynet}, G., \& {Maeder}, A. 2002, \aap, 386, 576

\bibitem[{{Eldridge} {et~al.}(2008){Eldridge}, {Izzard}, \&
  {Tout}}]{Eldridge2008}
{Eldridge}, J.~J., {Izzard}, R.~G., \& {Tout}, C.~A. 2008, \mnras, 384, 1109

\bibitem[{{Gazak} {et~al.}(2014){Gazak}, {Davies}, {Bastian}, {Kudritzki},
  {Bergemann}, {Plez}, {Evans}, {Patrick}, {Bresolin}, \&
  {Schinnerer}}]{Gazak2014}
{Gazak}, J.~Z., {Davies}, B., {Bastian}, N., {et~al.} 2014, \apj, 787, 142

\bibitem[{{Girardi} {et~al.}(2000){Girardi}, {Bressan}, {Bertelli}, \&
  {Chiosi}}]{Girardi2000}
{Girardi}, L., {Bressan}, A., {Bertelli}, G., \& {Chiosi}, C. 2000, VizieR
  Online Data Catalog, J/A+AS/141/371

\bibitem[{{Girardi} {et~al.}(2008){Girardi}, {Dalcanton}, {Williams}, {de
  Jong}, {Gallart}, {Monelli}, {Groenewegen}, {Holtzman}, {Olsen}, {Seth},
  {Weisz}, \& {ANGST/ANGRRR Collaboration}}]{Girardi2008}
{Girardi}, L., {Dalcanton}, J., {Williams}, B., {et~al.} 2008, \pasp, 120, 583

\bibitem[{{Gonz{\'a}lez Delgado} {et~al.}(1997){Gonz{\'a}lez Delgado},
  {Leitherer}, {Heckman}, \& {Cervi{\~n}o}}]{Gonzales1997}
{Gonz{\'a}lez Delgado}, R.~M., {Leitherer}, C., {Heckman}, T., \&
  {Cervi{\~n}o}, M. 1997, \apj, 483, 705

\bibitem[{{Gratton} {et~al.}(2004){Gratton}, {Sneden}, \&
  {Carretta}}]{Gratton2004}
{Gratton}, R., {Sneden}, C., \& {Carretta}, E. 2004, \araa, 42, 385

\bibitem[{{Greggio} {et~al.}(1998){Greggio}, {Tosi}, {Clampin}, {De Marchi},
  {Leitherer}, {Nota}, \& {Sirianni}}]{Greggio1998}
{Greggio}, L., {Tosi}, M., {Clampin}, M., {et~al.} 1998, \apj, 504, 725

\bibitem[{{Grocholski} {et~al.}(2008){Grocholski}, {Aloisi}, {van der Marel},
  {Mack}, {Annibali}, {Angeretti}, {Greggio}, {Held}, {Romano}, {Sirianni}, \&
  {Tosi}}]{Grocholski2008}
{Grocholski}, A.~J., {Aloisi}, A., {van der Marel}, R.~P., {et~al.} 2008,
  \apjl, 686, L79

\bibitem[{{Gustafsson} {et~al.}(2008){Gustafsson}, {Edvardsson}, {Eriksson},
  {J{\o}rgensen}, {Nordlund}, \& {Plez}}]{Gustafsson2008}
{Gustafsson}, B., {Edvardsson}, B., {Eriksson}, K., {et~al.} 2008, \aap, 486,
  951

\bibitem[{{Hagen} \& {van den Bergh}(1974)}]{Hagen1974}
{Hagen}, G.~L. \& {van den Bergh}, S. 1974, \apjl, 189, L103

\bibitem[{{Hernandez} {et~al.}(2017){Hernandez}, {Larsen}, {Trager}, {Groot},
  \& {Kaper}}]{Hernandez2017}
{Hernandez}, S., {Larsen}, S., {Trager}, S., {Groot}, P., \& {Kaper}, L. 2017,
  \aap, 603, A119

\bibitem[{{Herrero} {et~al.}(2020){Herrero}, {Parthasarathy},
  {Sim{\'o}n-D{\'\i}az}, {Hubrig}, {Sarkar}, \& {Muneer}}]{Herrero2020}
{Herrero}, A., {Parthasarathy}, M., {Sim{\'o}n-D{\'\i}az}, S., {et~al.} 2020,
  \mnras, 494, 2117

\bibitem[{{Hill}(1999)}]{Hill1999}
{Hill}, V. 1999, \aap, 345, 430

\bibitem[{{Hinkle} {et~al.}(2000){Hinkle}, {Wallace}, {Valenti}, \&
  {Harmer}}]{Wallace2000}
{Hinkle}, K., {Wallace}, L., {Valenti}, J., \& {Harmer}, D. 2000, {Visible and
  Near Infrared Atlas of the Arcturus Spectrum 3727-9300 A}

\bibitem[{{Ho} \& {Filippenko}(1996)}]{Ho1996}
{Ho}, L.~C. \& {Filippenko}, A.~V. 1996, \apjl, 466, L83

\bibitem[{{Humphreys} \& {McElroy}(1984)}]{Humphreys1984}
{Humphreys}, R.~M. \& {McElroy}, D.~B. 1984, \apj, 284, 565

\bibitem[{{Hunter} \& {Elmegreen}(2004)}]{Hunter2004}
{Hunter}, D.~A. \& {Elmegreen}, B.~G. 2004, \aj, 128, 2170

\bibitem[{{Israel}(1988)}]{Israel1988}
{Israel}, F.~P. 1988, \aap, 194, 24

\bibitem[{{Ivanov}(1998)}]{Ivanov1998}
{Ivanov}, G.~R. 1998, \aap, 337, 39

\bibitem[{{Kobayashi} {et~al.}(2020){Kobayashi}, {Karakas}, \&
  {Lugaro}}]{Kobayashi2020}
{Kobayashi}, C., {Karakas}, A.~I., \& {Lugaro}, M. 2020, \apj, 900, 179

\bibitem[{{Kraft} {et~al.}(1997){Kraft}, {Sneden}, {Smith}, {Shetrone},
  {Langer}, \& {Pilachowski}}]{Kraft1997}
{Kraft}, R.~P., {Sneden}, C., {Smith}, G.~H., {et~al.} 1997, \aj, 113, 279

\bibitem[{{Kudritzki} {et~al.}(2014){Kudritzki}, {Urbaneja}, {Bresolin},
  {Hosek}, \& {Przybilla}}]{Kudritzki2014}
{Kudritzki}, R.-P., {Urbaneja}, M.~A., {Bresolin}, F., {Hosek}, Matthew~W., J.,
  \& {Przybilla}, N. 2014, \apj, 788, 56

\bibitem[{{Kurucz}(1970)}]{Kurucz1970}
{Kurucz}, R.~L. 1970, SAO Special Report, 309

\bibitem[{{Kurucz} \& {Furenlid}(1979)}]{Kurucz1979}
{Kurucz}, R.~L. \& {Furenlid}, I. 1979, SAO Special Report, 387

\bibitem[{{Kurucz} {et~al.}(1984){Kurucz}, {Furenlid}, {Brault}, \&
  {Testerman}}]{Kurucz1984}
{Kurucz}, R.~L., {Furenlid}, I., {Brault}, J., \& {Testerman}, L. 1984, {Solar
  flux atlas from 296 to 1300 nm}

\bibitem[{{La Barbera} {et~al.}(2013){La Barbera}, {Ferreras}, {Vazdekis}, {de
  la Rosa}, {de Carvalho}, {Trevisan}, {Falc{\'o}n-Barroso}, \&
  {Ricciardelli}}]{La_Barbera2013}
{La Barbera}, F., {Ferreras}, I., {Vazdekis}, A., {et~al.} 2013, \mnras, 433,
  3017

\bibitem[{{Langer} \& {Maeder}(1995)}]{Langer1995}
{Langer}, N. \& {Maeder}, A. 1995, \aap, 295, 685

\bibitem[{{Lardo} {et~al.}(2017){Lardo}, {Cabrera-Ziri}, {Davies}, \&
  {Bastian}}]{Lardo2017}
{Lardo}, C., {Cabrera-Ziri}, I., {Davies}, B., \& {Bastian}, N. 2017, \mnras,
  468, 2482

\bibitem[{{Larsen} {et~al.}(2014){Larsen}, {Brodie}, {Forbes}, \&
  {Strader}}]{Larsen2014}
{Larsen}, S.~S., {Brodie}, J.~P., {Forbes}, D.~A., \& {Strader}, J. 2014, \aap,
  565, A98

\bibitem[{{Larsen} {et~al.}(2012){Larsen}, {Brodie}, \& {Strader}}]{Larsen2012}
{Larsen}, S.~S., {Brodie}, J.~P., \& {Strader}, J. 2012, \aap, 546, A53

\bibitem[{{Larsen} {et~al.}(2011){Larsen}, {de Mink}, {Eldridge}, {Langer},
  {Bastian}, {Seth}, {Smith}, {Brodie}, \& {Efremov}}]{Larsen2011}
{Larsen}, S.~S., {de Mink}, S.~E., {Eldridge}, J.~J., {et~al.} 2011, \aap, 532,
  A147

\bibitem[{{Larsen} {et~al.}(2022){Larsen}, {Eitner}, {Magg}, {Bergemann},
  {Moltzer}, {Brodie}, {Romanowsky}, \& {Strader}}]{Larsen2022}
{Larsen}, S.~S., {Eitner}, P., {Magg}, E., {et~al.} 2022, arXiv e-prints,
  arXiv:2112.00081

\bibitem[{{Larsen} {et~al.}(2008){Larsen}, {Origlia}, {Brodie}, \&
  {Gallagher}}]{Larsen2008}
{Larsen}, S.~S., {Origlia}, L., {Brodie}, J., \& {Gallagher}, J.~S. 2008,
  \mnras, 383, 263

\bibitem[{{Larsen} {et~al.}(2006){Larsen}, {Origlia}, {Brodie}, \&
  {Gallagher}}]{Larsen2006}
{Larsen}, S.~S., {Origlia}, L., {Brodie}, J.~P., \& {Gallagher}, J.~S. 2006,
  \mnras, 368, L10

\bibitem[{Letarte(2007)}]{Letarte2007}
Letarte, B. 2007, PhD thesis, University of Groningen, date\_submitted:2007
  Rights: University of Groningen

\bibitem[{{Letarte} {et~al.}(2010){Letarte}, {Hill}, {Tolstoy}, {Jablonka},
  {Shetrone}, {Venn}, {Spite}, {Irwin}, {Battaglia}, {Helmi}, {Primas},
  {Fran{\c{c}}ois}, {Kaufer}, {Szeifert}, {Arimoto}, \&
  {Sadakane}}]{Letarte2010}
{Letarte}, B., {Hill}, V., {Tolstoy}, E., {et~al.} 2010, \aap, 523, A17

\bibitem[{{Lyubimkov} {et~al.}(2004){Lyubimkov}, {Rostopchin}, \&
  {Lambert}}]{Lyubimkov2004}
{Lyubimkov}, L.~S., {Rostopchin}, S.~I., \& {Lambert}, D.~L. 2004, \mnras, 351,
  745

\bibitem[{{Marigo} {et~al.}(2008){Marigo}, {Girardi}, {Bressan}, {Groenewegen},
  {Silva}, \& {Granato}}]{Marigo2008}
{Marigo}, P., {Girardi}, L., {Bressan}, A., {et~al.} 2008, \aap, 482, 883

\bibitem[{{Martocchia} {et~al.}(2018){Martocchia}, {Cabrera-Ziri}, {Lardo},
  {Dalessandro}, {Bastian}, {Kozhurina-Platais}, {Usher}, {Niederhofer},
  {Cordero}, {Geisler}, {Hollyhead}, {Kacharov}, {Larsen}, {Li}, {Mackey},
  {Hilker}, {Mucciarelli}, {Platais}, \& {Salaris}}]{Martocchia2018}
{Martocchia}, S., {Cabrera-Ziri}, I., {Lardo}, C., {et~al.} 2018, \mnras, 473,
  2688

\bibitem[{{Martocchia} {et~al.}(2019){Martocchia}, {Dalessandro}, {Lardo},
  {Cabrera-Ziri}, {Bastian}, {Kozhurina-Platais}, {Salaris}, {Chantereau},
  {Geisler}, {Hilker}, {Kacharov}, {Larsen}, {Mucciarelli}, {Niederhofer},
  {Platais}, \& {Usher}}]{Martocchia2019}
{Martocchia}, S., {Dalessandro}, E., {Lardo}, C., {et~al.} 2019, \mnras, 487,
  5324

\bibitem[{{McWilliam}(1997)}]{McWilliam1997}
{McWilliam}, A. 1997, \araa, 35, 503

\bibitem[{{Mengel} {et~al.}(1979){Mengel}, {Sweigart}, {Demarque}, \&
  {Gross}}]{Mengel1979}
{Mengel}, J.~G., {Sweigart}, A.~V., {Demarque}, P., \& {Gross}, P.~G. 1979,
  \apjs, 40, 733

\bibitem[{{Meylan} \& {Maeder}(1982)}]{Meylan1982}
{Meylan}, G. \& {Maeder}, A. 1982, \aap, 108, 148

\bibitem[{{Minelli} {et~al.}(2021){Minelli}, {Mucciarelli}, {Romano},
  {Bellazzini}, {Origlia}, \& {Ferraro}}]{Minelli2021}
{Minelli}, A., {Mucciarelli}, A., {Romano}, D., {et~al.} 2021, \apj, 910, 114

\bibitem[{{Pagel}(2009)}]{Pagel2009book}
{Pagel}, B. E.~J. 2009, {Nucleosynthesis and Chemical Evolution of Galaxies}

\bibitem[{{Plez}(2012)}]{Plez2012}
{Plez}, B. 2012, {Turbospectrum: Code for spectral synthesis}

\bibitem[{{Reddy} {et~al.}(2006){Reddy}, {Lambert}, \& {Allende
  Prieto}}]{Reddy2006}
{Reddy}, B.~E., {Lambert}, D.~L., \& {Allende Prieto}, C. 2006, \mnras, 367,
  1329

\bibitem[{{Reddy} {et~al.}(2003){Reddy}, {Tomkin}, {Lambert}, \& {Allende
  Prieto}}]{Reddy2003}
{Reddy}, B.~E., {Tomkin}, J., {Lambert}, D.~L., \& {Allende Prieto}, C. 2003,
  \mnras, 340, 304

\bibitem[{{Robertson}(1973)}]{Robertson1973}
{Robertson}, J.~W. 1973, \apj, 180, 425

\bibitem[{{Robertson}(1974)}]{Robertson1974}
{Robertson}, J.~W. 1974, \apj, 191, 67

\bibitem[{{Romano} {et~al.}(2006){Romano}, {Tosi}, \& {Matteucci}}]{Romano2006}
{Romano}, D., {Tosi}, M., \& {Matteucci}, F. 2006, \mnras, 365, 759

\bibitem[{{Sakari} {et~al.}(2021){Sakari}, {Shetrone}, {McWilliam}, \&
  {Wallerstein}}]{Sakari2021}
{Sakari}, C.~M., {Shetrone}, M.~D., {McWilliam}, A., \& {Wallerstein}, G. 2021,
  \mnras, 502, 5745

\bibitem[{{Salpeter}(1955)}]{Salpeter1955}
{Salpeter}, E.~E. 1955, \apj, 121, 161

\bibitem[{{Salvador-Rusi{\~n}ol} {et~al.}(2021){Salvador-Rusi{\~n}ol},
  {Beasley}, {Vazdekis}, \& {Barbera}}]{Nuria2021}
{Salvador-Rusi{\~n}ol}, N., {Beasley}, M.~A., {Vazdekis}, A., \& {Barbera},
  F.~L. 2021, \mnras, 500, 3368

\bibitem[{{Shetrone}(1996)}]{Shetrone1996}
{Shetrone}, M.~D. 1996, \aj, 112, 2639

\bibitem[{{Shields}(1990)}]{Shields1990}
{Shields}, G.~A. 1990, \araa, 28, 525

\bibitem[{{Sternberg}(1998)}]{Sternberg1998}
{Sternberg}, A. 1998, \apj, 506, 721

\bibitem[{{Stil} \& {Israel}(2002)}]{Stil_Israel2002}
{Stil}, J.~M. \& {Israel}, F.~P. 2002, \aap, 392, 473

\bibitem[{{Tolstoy} {et~al.}(2009){Tolstoy}, {Hill}, \& {Tosi}}]{Tolstoy2009}
{Tolstoy}, E., {Hill}, V., \& {Tosi}, M. 2009, \araa, 47, 371

\bibitem[{{Truran}(1981)}]{Truran1981}
{Truran}, J.~W. 1981, \aap, 97, 391

\bibitem[{{Vallenari} \& {Bomans}(1996)}]{Vallenari1996}
{Vallenari}, A. \& {Bomans}, D.~J. 1996, \aap, 313, 713

\bibitem[{{van den Bergh}(1968)}]{Bergh1968}
{van den Bergh}, S. 1968, \jrasc, 62, 219

\bibitem[{{Van der Swaelmen} {et~al.}(2013){Van der Swaelmen}, {Hill},
  {Primas}, \& {Cole}}]{VanderSwaelmen2013}
{Van der Swaelmen}, M., {Hill}, V., {Primas}, F., \& {Cole}, A.~A. 2013, \aap,
  560, A44

\end{thebibliography}

\appendix
\onecolumn
\section{HRD FILE}

\begin{longtable}{cccc|ccc|cc}
\caption{HRD file}
\label{Tab_hrd}\\
\hline
\hline
MASS & TEFF & LOGG & RSTAR & & WEIGHT & & LOGVT & SYNT  \\
 &  &  &  & $N_{RSG}/N_{BSG}$=1.24 & $N_{RSG}/N_{BSG}$=1.53 & $N_{RSG}/N_{BSG}$=1.90 &  &    \\
\hline
\endfirsthead
\caption{Continued. HRD file} \\
\hline
\hline
MASS & TEFF & LOGG & RSTAR & & WEIGHT & & LOGVT & SYNT  \\
 &  &  &  & $N_{RSG}/N_{BSG}$=1.24 & $N_{RSG}/N_{BSG}$=1.53 & $N_{RSG}/N_{BSG}$=1.90 &  &   \\
\hline
\endhead
\hline
\endfoot
\hline
\endlastfoot
0.250 & 3742.8 & 5.0615 & 0.264 & 2.599e+00 & 2.599e+00 & 2.599e+00 & 0.301 & MT \\
0.350 & 3892.2 & 4.9995 & 0.333 & 1.179e+00 & 1.179e+00 & 1.179e+00 & 0.301 & MT \\
0.450 & 4037.4 & 4.9064 & 0.412 & 6.531e-01 & 6.531e-01 & 6.531e-01 & 0.301 & A9S \\
0.550 & 4257.0 & 4.8160 & 0.497 & 4.075e-01 & 4.075e-01 & 4.075e-01 & 0.301 & A9S \\
0.700 & 4719.5 & 4.6872 & 0.658 & 3.468e-01 & 3.468e-01 & 3.468e-01 & 0.301 & A9S \\
0.900 & 5679.4 & 4.5922 & 0.812 & 2.562e-01 & 2.562e-01 & 2.562e-01 & 0.301 & A9S \\
1.100 & 6376.8 & 4.4495 & 1.040 & 1.599e-01 & 1.599e-01 & 1.599e-01 & 0.602 & A9S \\
1.300 & 7074.6 & 4.3546 & 1.245 & 1.080e-01 & 1.080e-01 & 1.080e-01 & 0.602 & A9S \\
1.500 & 8087.2 & 4.3711 & 1.310 & 7.713e-02 & 7.713e-02 & 7.713e-02 & 0.602 & A9S \\
1.700 & 9084.5 & 4.3967 & 1.355 & 5.747e-02 & 5.747e-02 & 5.747e-02 & 0.602 & A9S \\
1.900 & 9954.1 & 4.4096 & 1.397 & 4.425e-02 & 4.425e-02 & 4.425e-02 & 0.602 & A9S \\
2.200 & 11043.3 & 4.4087 & 1.465 & 4.704e-02 & 4.704e-02 & 4.704e-02 & 0.602 & A9S \\
3.000 & 13418.4 & 4.3808 & 1.708 & 6.051e-02 & 6.051e-02 & 6.051e-02 & 0.602 & A9S \\
3.500 & 14608.3 & 4.3537 & 1.892 & 2.633e-02 & 2.633e-02 & 2.633e-02 & 0.602 & A9S \\
4.500 & 16795.8 & 4.3123 & 2.211 & 2.917e-02 & 2.917e-02 & 2.917e-02 & 0.602 & A9S \\
5.000 & 17762.3 & 4.2852 & 2.386 & 1.139e-02 & 1.139e-02 & 1.139e-02 & 0.602 & A9S \\
6.000 & 19494.0 & 4.2216 & 2.772 & 1.484e-02 & 1.484e-02 & 1.484e-02 & 0.602 & A9S \\
7.000 & 20979.7 & 4.1437 & 3.225 & 1.033e-02 & 1.033e-02 & 1.033e-02 & 0.602 & A9S \\
7.960 & 21973.5 & 4.0359 & 3.772 & 7.330e-03 & 7.330e-03 & 7.330e-03 & 0.602 & A9S \\
8.783 & 22356.3 & 3.8979 & 4.653 & 4.990e-03 & 4.990e-03 & 4.990e-03 & 0.903 & A9S \\
9.094 & 22371.8 & 3.8351 & 5.102 & 1.733e-03 & 1.733e-03 & 1.733e-03 & 0.903 & A9S \\
10.119 & 20682.4 & 3.4763 & 8.149 & 4.454e-03 & 4.454e-03 & 4.454e-03 & 0.602 & A9S \\
10.375 & 19865.5 & 3.3399 & 9.883 & 1.048e-03 & 1.048e-03 & 1.048e-03 & 0.602 & A9S \\
10.443 & 21488.2 & 3.4329 & 8.681 & 2.764e-04 & 2.764e-04 & 2.764e-04 & 0.602 & A9S \\
10.447 & 22610.0 & 3.4851 & 8.031 & 1.358e-05 & 1.358e-05 & 1.358e-05 & 0.903 & A9S \\
10.448 & 20936.3 & 3.3937 & 9.184 & 6.971e-06 & 6.971e-06 & 6.971e-06 & 0.602 & A9S \\
10.449 & 19257.5 & 3.2362 & 11.377 & 3.873e-06 & 3.873e-06 & 3.873e-06 & 0.602 & A9S \\
10.450 & 17910.2 & 3.0969 & 13.134 & 2.989e-06 & 2.989e-06 & 2.989e-06 & 0.602 & A9S \\
10.451 & 16421.0 & 2.9354 & 16.494 & 3.122e-06 & 3.122e-06 & 3.122e-06 & 0.602 & A9S \\
10.452 & 14746.9 & 2.7428 & 21.223 & 3.391e-06 & 3.391e-06 & 3.391e-06 & 0.602 & A9S \\
10.452 & 13329.1 & 2.5668 & 25.592 & 2.684e-06 & 2.684e-06 & 2.684e-06 & 0.602 & A9S \\
10.453 & 12075.4 & 2.3982 & 31.745 & 2.188e-06 & 2.188e-06 & 2.188e-06 & 0.602 & A9S \\
10.453 & 11043.3 & 2.2477 & 39.213 & 1.697e-06 & 1.697e-06 & 1.697e-06 & 0.602 & A9S \\
10.454 & 9988.5 & 2.0805 & 45.618 & 1.554e-06 & 1.554e-06 & 1.554e-06 & 0.602 & A9S \\
10.454 & 9017.8 & 1.9116 & 59.362 & 1.339e-06 & 1.339e-06 & 1.339e-06 & 0.602 & A9S \\
10.454 & 8352.2 & 1.7860 & 69.308 & 9.172e-07 & 9.172e-07 & 9.172e-07 & 0.602 & A9S \\
10.454 & 7622.5 & 1.6377 & 82.409 & 9.210e-07 & 9.210e-07 & 9.210e-07 & 0.602 & A9S \\
10.455 & 6935.9 & 1.4865 & 97.593 & 8.442e-07 & 8.442e-07 & 8.442e-07 & 0.602 & A9S \\
10.455 & 6181.6 & 1.3067 & 119.795 & 8.480e-07 & 8.480e-07 & 8.480e-07 & 0.602 & A9S \\
10.455 & 5393.9 & 1.1062 & 153.595 & 8.480e-07 & 7.517e-07 & 6.558e-07 & 0.301 & A9S \\
10.455 & 4842.8 & 0.9659 & 185.366 & 6.331e-07 & 6.909e-07 & 7.485e-07 & 0.301 & A9S \\
10.455 & 4478.2 & 0.8931 & 204.328 & 9.898e-07 & 1.080e-06 & 1.170e-06 & 0.301 & A9S \\
10.456 & 4272.7 & 0.7605 & 235.374 & 7.865e-07 & 8.583e-07 & 9.299e-07 & 0.301 & A9S \\
10.456 & 4066.3 & 0.4630 & 336.095 & 9.591e-07 & 1.047e-06 & 1.134e-06 & 0.301 & A9S \\
10.456 & 4004.1 & 0.3326 & 390.427 & 1.661e-06 & 1.813e-06 & 1.964e-06 & 0.301 & A9S \\
10.459 & 3944.6 & 0.2259 & 436.590 & 9.516e-06 & 1.038e-05 & 1.125e-05 & 0.301 & MT \\
10.489 & 3980.2 & 0.2752 & 410.120 & 1.213e-04 & 1.324e-04 & 1.435e-04 & 0.301 & MT \\
10.538 & 4077.6 & 0.4221 & 354.545 & 1.917e-04 & 2.093e-04 & 2.267e-04 & 0.301 & A9S \\
10.611 & 4220.9 & 0.5417 & 306.315 & 2.848e-04 & 3.108e-04 & 3.367e-04 & 0.301 & A9S \\
10.640 & 4462.7 & 0.5851 & 291.341 & 1.136e-04 & 1.240e-04 & 1.343e-04 & 0.301 & A9S \\
10.644 & 5862.7 & 1.0110 & 171.486 & 1.513e-05 & 1.341e-05 & 1.170e-05 & 0.301 & A9S \\
10.646 & 6858.0 & 1.2818 & 123.846 & 7.295e-06 & 6.467e-06 & 5.642e-06 & 0.602 & A9S \\
10.649 & 7585.8 & 1.4575 & 101.760 & 9.522e-06 & 8.441e-06 & 7.364e-06 & 0.602 & A9S \\
10.655 & 8369.5 & 1.6262 & 83.102 & 2.399e-05 & 2.126e-05 & 1.855e-05 & 0.602 & A9S \\
10.664 & 9046.9 & 1.7554 & 73.575 & 3.704e-05 & 3.283e-05 & 2.864e-05 & 0.602 & A9S \\
10.677 & 9772.4 & 1.8828 & 59.004 & 4.838e-05 & 4.289e-05 & 3.741e-05 & 0.602 & A9S \\
10.694 & 10497.8 & 2.0002 & 50.750 & 6.432e-05 & 5.702e-05 & 4.974e-05 & 0.602 & A9S \\
10.723 & 11071.3 & 2.0840 & 45.245 & 1.098e-04 & 9.737e-05 & 8.495e-05 & 0.602 & A9S \\
10.764 & 10136.8 & 1.9261 & 55.671 & 1.533e-04 & 1.359e-04 & 1.186e-04 & 0.602 & A9S \\
10.780 & 9313.2 & 1.7798 & 68.492 & 5.974e-05 & 5.296e-05 & 4.620e-05 & 0.602 & A9S \\
10.791 & 8480.1 & 1.6190 & 82.935 & 4.138e-05 & 3.668e-05 & 3.200e-05 & 0.602 & A9S \\
10.798 & 7730.4 & 1.4618 & 102.336 & 2.766e-05 & 2.452e-05 & 2.139e-05 & 0.602 & A9S \\
10.803 & 7104.0 & 1.3192 & 119.728 & 1.852e-05 & 1.641e-05 & 1.432e-05 & 0.602 & A9S \\
10.807 & 6478.9 & 1.1655 & 144.114 & 1.433e-05 & 1.270e-05 & 1.108e-05 & 0.602 & A9S \\
10.810 & 5833.1 & 0.9920 & 176.415 & 1.101e-05 & 9.764e-06 & 8.518e-06 & 0.301 & A9S \\
10.812 & 5311.3 & 0.8432 & 213.002 & 6.727e-06 & 5.963e-06 & 5.202e-06 & 0.301 & A9S \\
10.815 & 4305.3 & 0.5574 & 304.855 & 1.287e-05 & 1.404e-05 & 1.521e-05 & 0.301 & A9S \\
10.820 & 4102.0 & 0.4059 & 366.941 & 1.900e-05 & 2.074e-05 & 2.247e-05 & 0.301 & A9S \\
10.822 & 4080.4 & 0.3882 & 374.272 & 6.567e-06 & 7.167e-06 & 7.764e-06 & 0.301 & A9S \\
10.825 & 3994.8 & 0.2721 & 418.586 & 1.138e-05 & 1.242e-05 & 1.346e-05 & 0.301 & MT \\
10.829 & 3879.7 & 0.0817 & 533.224 & 1.528e-05 & 1.668e-05 & 1.807e-05 & 0.301 & MT \\
10.834 & 3769.6 & -0.0637 & 630.534 & 1.552e-05 & 1.694e-05 & 1.835e-05 & 0.301 & MT \\

\end{longtable}

\clearpage
\onecolumn
\section{Chemical abundances}

\begin{longtable}{cc|ccc}
\caption{Chemical abundances for NGC 1569-B}
\label{Tab_abun}\\
\hline
\hline
Element & Wavelength ({\AA}) & & Abundance &  \\
 &  &   $N_{RSG}/N_{BSG}$=1.24 & $N_{RSG}/N_{BSG}$=1.53 & $N_{RSG}/N_{BSG}$=1.90  \\
\hline
\endfirsthead
\caption{Continued. Chemical abundances for NGC 1569-B} \\
\hline
\hline
Element & Wavelength ({\AA}) & & Abundance &  \\
 &  &   $N_{RSG}/N_{BSG}$=1.24 & $N_{RSG}/N_{BSG}$=1.53 & $N_{RSG}/N_{BSG}$=1.90  \\
\hline
\endhead
\hline
\endfoot
\hline
\endlastfoot
[Fe/H]  &       4170.0-4180.0   &       -0.243  $\pm$   0.058   &       -               &       -       \\
        &       4232.4-4240.0   &       -0.253  $\pm$   0.050   &       -               &       -               \\
        &       4293.0-4316.0   &       -0.403  $\pm$   0.041   &       -0.350  $\pm$   0.038   &       -0.299  $\pm$   0.034   \\
        &       4400.0-4424.0 & -1.003  $\pm$   0.045   &       -1.055  $\pm$   0.048   &       -1.095  $\pm$   0.051   \\
        &       4500.0-4530.0 & -0.786  $\pm$   0.042   &       -0.853  $\pm$   0.047   &       -0.916  $\pm$   0.046   \\
        &       4573.0-4600.0 & -0.828  $\pm$   0.045   &       -0.964  $\pm$   0.045   &       -1.042  $\pm$   0.044   \\
        &       4631.0-4660.0 & -0.887  $\pm$   0.060   &       -0.896  $\pm$   0.068   &       -1.007  $\pm$   0.078   \\
        &       4706.0-4711.0 & -1.007  $\pm$   0.127   &       -1.116  $\pm$   0.126   &       -1.216  $\pm$   0.119   \\
        &       4724.0-4750.0 & -0.651  $\pm$   0.057   &       -0.797  $\pm$   0.054   &       -0.873  $\pm$   0.052   \\
        &       4869.0-4883.0 & -0.568  $\pm$   0.049   &       -0.639  $\pm$   0.052   &       -0.718  $\pm$   0.053   \\
        &       4897.0-4915.0 & -1.017  $\pm$   0.057   &       -1.103  $\pm$   0.057   &       -1.142  $\pm$   0.056   \\
        &       4918.0-4926.0 & 0.101   $\pm$   0.057   &       -               &       -               \\
        &       4936.0-4944.0 & -0.513  $\pm$   0.082   &       -0.667  $\pm$   0.087   &       -0.765  $\pm$   0.089   \\
        &       4945.5-4953.0 & -0.338  $\pm$   0.132   &       -0.455  $\pm$   0.137   &       -0.440  $\pm$   0.165   \\
        &       4963.0-4976.0 & -1.361  $\pm$   0.091   &       -       &       -       \\
        &       5122.4-5150.0 & -0.861  $\pm$   0.036   &       -1.00 $\pm$     0.037   &       -1.132  $\pm$   0.036   \\
        &       5224.0-5235.0 & -0.159  $\pm$   0.037   &       -0.272  $\pm$   0.036   &       -0.293  $\pm$   0.049   \\
        &       5266.0-5289.0 & -0.481  $\pm$   0.022   &       -0.517  $\pm$   0.024   &       -0.536  $\pm$   0.027   \\
        &       5250.0-5259.0 & -0.699  $\pm$   0.054   &       -0.784  $\pm$   0.056   &       -0.813  $\pm$   0.059   \\
        &       5300.0-5345.0 & -0.833  $\pm$   0.021   &       -0.923  $\pm$   0.022   &       -1.007  $\pm$   0.022   \\
        &       5358.0-5375.0 & -0.216  $\pm$   0.022   &       -0.248  $\pm$   0.021   &       -0.255  $\pm$   0.022   \\
        &       5378.0-5400.0 & -0.686  $\pm$   0.024   &       -0.769  $\pm$   0.026   &       -0.811  $\pm$   0.025   \\
        &       5400.0-5420.0 & -0.542  $\pm$   0.025   &       -0.599  $\pm$   0.025   &       -0.654  $\pm$   0.025   \\
        &       5420.0-5449.0 & -0.236  $\pm$   0.016   &       -0.270  $\pm$   0.019   &       -0.257  $\pm$   0.018   \\
        &       5460.0-5475.5   &       -1.022  $\pm$   0.057   &       -1.167  $\pm$   0.056   &       -1.285  $\pm$   0.051   \\
        &       5494.0-5510.0 & -0.598  $\pm$   0.053   &       -0.791  $\pm$   0.048   &       -0.792  $\pm$   0.046   \\
        &       5529.0-5539.0 & -0.458  $\pm$   0.040   &       -0.557  $\pm$   0.038   &       -0.598  $\pm$   0.037   \\
        &       5566.5-5590.0 & -0.598  $\pm$   0.046   &       -0.801  $\pm$   0.041   &       -0.896  $\pm$   0.038   \\
        &       5610.2-5630.0 & -0.390  $\pm$   0.052   &       -0.568  $\pm$   0.047   &       -0.695  $\pm$   0.044   \\
        &       5682.0-5708.0 & -0.842  $\pm$   0.037   &       -0.872  $\pm$   0.035   &       -0.903  $\pm$   0.034   \\
        &       5708.0-5714.0 & -0.768  $\pm$   0.058   &       -0.793  $\pm$   0.058   &       -0.789  $\pm$   0.059   \\
        &       5858.5-5865.0 & -0.717  $\pm$   0.095   &       -0.781  $\pm$   0.097   &       -0.847  $\pm$   0.101   \\
        &       5970.0-5980.0 & -0.707  $\pm$   0.083   &       -0.854  $\pm$   0.077   &       -0.937  $\pm$   0.070   \\
        &       6001.0-6030.0 & -0.682  $\pm$   0.035   &       -0.776  $\pm$   0.033   &       -0.839  $\pm$   0.031   \\
        &       6053.0-6082.0 & -0.574  $\pm$   0.038   &       -0.700  $\pm$   0.037   &       -0.779  $\pm$   0.034   \\
        &       6131.0-6140.0 & -0.220  $\pm$   0.031   &       -0.335  $\pm$   0.031   &       -0.409  $\pm$   0.032   \\
        &       6144.0-6160.0 & -0.309  $\pm$   0.032   &       -0.371  $\pm$   0.034   &       -0.386  $\pm$   0.035   \\
        &       6170.0-6185.0 & -0.651  $\pm$   0.056   &       -0.777  $\pm$   0.060   &       -0.853  $\pm$   0.059   \\
        
  &  &   &  & \\
  
[Mg/Fe] &       4347.0-4357.0 & -0.004  $\pm$   0.111   &       0.162   $\pm$   0.087   &       0.361   $\pm$   0.068   \\
        &       4565.0-4576.0 & -0.184  $\pm$   0.113   &       0.017   $\pm$   0.091   &       0.219   $\pm$   0.072   \\
        &       4700.0-4707.0 & 0.313   $\pm$   0.117   &       0.487   $\pm$   0.125   &       0.365   $\pm$   0.120   \\
        &       5160.0-5190.0 & 0.065   $\pm$   0.020   &       0.133   $\pm$   0.019   &       0.236   $\pm$   0.022   \\
        &       5523.0-5531.5   &       0.327   $\pm$   0.050   &       0.300   $\pm$   0.042   &       0.274   $\pm$   0.040   \\
        &       5705.0-5715.0 & -0.317  $\pm$   0.054   &       -0.135  $\pm$   0.050   &       -0.014  $\pm$   0.048   \\
        
  &  &   &  & \\                
  
[Ca/Fe] &       4220.0-4234.0 & -0.596  $\pm$   0.073   &       -0.617  $\pm$   0.066   &       -0.673  $\pm$   0.064   \\
        &       4420.0-4440.0 & 0.811   $\pm$   0.074   &       0.943   $\pm$   0.079   &       0.753   $\pm$   0.090   \\
        &       4451.0-4461.0 & 0.132   $\pm$   0.148   &       0.113   $\pm$   0.177   &       -0.192  $\pm$   0.168   \\
        &       4521.0-4531.0 & -0.49   $\pm$   0.274   &       -0.639  $\pm$   0.254   &       -0.643  $\pm$   0.231   \\
        &       4573.0-4590.0 & -0.981  $\pm$   0.149   &       -1.123  $\pm$   0.142   &       -1.226  $\pm$   0.140   \\
        &       5256.0-5268.0 & 0.664   $\pm$   0.097   &       0.573   $\pm$   0.124   &       0.425   $\pm$   0.112   \\
        &       5347.0-5357.0 & 0.364   $\pm$   0.124   &       0.351   $\pm$   0.134   &       0.280   $\pm$   0.135   \\
        &       5507.0-5517.0 & -0.484  $\pm$   0.149   &       -0.560  $\pm$   0.130   &       -0.573  $\pm$   0.114   \\
        &       5576.0-5602.0 & 0.557   $\pm$   0.050   &       0.442   $\pm$   0.058   &       0.209   $\pm$   0.052   \\
        &       5852.0-5862.0 & 0.532   $\pm$   0.055   &       0.489   $\pm$   0.059   &       0.402   $\pm$   0.059   \\
        &       6098.0-6127.0 & 0.844   $\pm$   0.021   &       0.823   $\pm$   0.022   &       0.722   $\pm$   0.024   \\
        &       6151.0-6174.0 & 0.366   $\pm$   0.034   &       0.185   $\pm$   0.038   &       -0.009  $\pm$   0.036   \\
          &  &   &  & \\
          
[Sc/Fe] &       4665.0-4675.0 & 0.520   $\pm$   0.165   &       0.487   $\pm$   0.188   &       0.381   $\pm$   0.189 \\
        &       4739.0-4758.0 & -0.236  $\pm$   0.191   &       -0.135  $\pm$   0.163   &       -0.076  $\pm$   0.148 \\
        &       5522.5-5531.0 & 1.127   $\pm$   0.059   &       1.202   $\pm$   0.059   &       1.191   $\pm$   0.058 \\
        &       5638.0-5690.0 & 1.049   $\pm$   0.036   &       0.949   $\pm$   0.041   &       0.833   $\pm$   0.040 \\
        &       6206.0-6216.0 & 0.588   $\pm$   0.171   &       0.328   $\pm$   0.123   &       0.243   $\pm$   0.106 \\
        
  &  &   &  & \\
  
[Ti/Fe] &       4293.0-4315.0 & 0.210   $\pm$   0.054   &       0.444   $\pm$   0.056   &       0.445   $\pm$   0.059   \\
        &       4442.0-4475.0 & 0.521   $\pm$   0.030   &       0.622   $\pm$   0.028   &       0.622   $\pm$   0.028   \\
        &       4521.0-4540.0 & 0.519   $\pm$   0.043   &       0.598   $\pm$   0.042   &       0.541   $\pm$   0.042   \\
        &       4570.0-4575.0 & 0.651   $\pm$   0.063   &       0.731   $\pm$   0.062   &       0.693   $\pm$   0.062   \\
        &       4500.0-4519.5   &       0.756   $\pm$   0.050   &       0.813   $\pm$   0.051   &       0.715   $\pm$   0.051   \\
        &       4551.0-4570.0 & 0.671   $\pm$   0.040   &       0.752   $\pm$   0.038   &       0.731   $\pm$   0.037   \\
        &       4638.0-4660.0 & 0.857   $\pm$   0.089   &       0.665   $\pm$   0.101   &       0.512   $\pm$   0.094   \\
        &       4680.0-4698.0 & 0.555   $\pm$   0.129   &       0.460   $\pm$   0.108   &       0.399   $\pm$   0.106   \\
        &       4802.0-4821.0 & 0.785   $\pm$   0.109   &       0.548   $\pm$   0.087   &       0.459   $\pm$   0.074   \\
        &       4975.0-4995.1   &       0.359   $\pm$   0.053   &       0.265   $\pm$   0.056   &       0.191   $\pm$   0.055   \\
        &       5510.0-5520.0 & 0.593   $\pm$   0.092   &       0.388   $\pm$   0.083   &       0.260   $\pm$   0.075   \\
        &       5860.0-5875.0 & 0.362   $\pm$   0.073   &       0.343   $\pm$   0.070   &       0.360   $\pm$   0.070   \\
        &       5912.0-5922.0 & 0.326   $\pm$   0.077   &       0.256   $\pm$   0.070   &       0.211   $\pm$   0.067   \\
        &       5960.0-5994.0   &       0.269   $\pm$   0.045   &       0.301   $\pm$   0.042   &       0.304   $\pm$   0.042   \\
        
  &  &   &  & \\
  
[Cr/Fe] &       4253.0-4260.0 & 0.872   $\pm$   0.110   &       1.260   $\pm$   0.093   &       1.417   $\pm$   0.086   \\
        &       4270.0-4276.0 & 0.589   $\pm$   0.101   &       0.306   $\pm$   0.107   &       0.384   $\pm$   0.107   \\
        &       4565.0-4570.0 & -0.284  $\pm$   0.285   &       -0.551  $\pm$   0.272   &       -0.683  $\pm$   0.285   \\
        &       4578.0-4597.0 & -0.046  $\pm$   0.052   &       0.051   $\pm$   0.052   &       0.100   $\pm$   0.051   \\
        &       4537.0-4550.0 & -0.265  $\pm$   0.195   &       -0.376  $\pm$   0.198   &       -0.430  $\pm$   0.187   \\
        &       4612.0-4631.0 & 0.281   $\pm$   0.057   &       0.326   $\pm$   0.064   &       0.314   $\pm$   0.065   \\
        &       4646.0-4657.0 & -0.760  $\pm$   0.190   &       -0.782  $\pm$   0.172   &       -0.790  $\pm$   0.161   \\
        &       4703.0-4723.0 & 0.406   $\pm$   0.111   &       0.314   $\pm$   0.110   &       0.260   $\pm$   0.103   \\
        &       4751.0-4761.0 & -0.075  $\pm$   0.171   &       0.028   $\pm$   0.153   &       0.102   $\pm$   0.143   \\
        &       4796.0-4806.0 & -0.240  $\pm$   0.374   &       -0.190  $\pm$   0.366   &       -0.318  $\pm$   0.286   \\
        &       4824.0-4834.0 & 0.476   $\pm$   0.087   &       0.589   $\pm$   0.091   &       0.566   $\pm$   0.092   \\
        &       4931.0-4947.0 & -1.029  $\pm$   0.258   &       -0.855  $\pm$   0.212   &       -0.717  $\pm$   0.195   \\
        &       5117.0-5127.0 & 1.551   $\pm$   0.133   &       1.516   $\pm$   0.127   &       1.532   $\pm$   0.113   \\
        &       5270.0-5281.0 & 0.194   $\pm$   0.070   &       0.199   $\pm$   0.067   &       0.238   $\pm$   0.063   \\
        &       5292.0-5304.0 & 0.500   $\pm$   0.047   &       0.193   $\pm$   0.066   &       -0.148  $\pm$   0.067   \\
        &       5341.0-5353.0 & 0.693   $\pm$   0.037   &       0.667   $\pm$   0.041   &       0.619   $\pm$   0.042   \\
        &       5407.0-5413.0 & 1.110   $\pm$   0.042   &       1.050   $\pm$   0.041   &       0.950   $\pm$   0.043   \\
        &       5779.0-5793.0 & 0.781   $\pm$   0.057   &       0.631   $\pm$   0.049   &       0.561   $\pm$   0.045   \\
        &       6325.0-6335.0 & 1.209   $\pm$   0.101   &       0.938   $\pm$   0.131   &       0.700   $\pm$   0.126   \\
        &       6973.0-6983.0 & 1.420   $\pm$   0.052   &       1.106   $\pm$   0.058   &       0.967   $\pm$   0.052   \\
        
  &  &   &  & \\
  
[Mn/Fe] &       5372.0-5382.0 & -0.874  $\pm$   0.209   &       -0.765  $\pm$   0.181   &       -0.723  $\pm$   0.171   \\
        &       5390.0-5410.0 & -0.124  $\pm$   0.080   &       -0.167  $\pm$   0.062   &       -0.159  $\pm$   0.057   \\
        &       5418.0-5434.0 & 0.395   $\pm$   0.104   &       0.139   $\pm$   0.104   &       0.031   $\pm$   0.095   \\
        &       5468.0-5490.0 & -0.460  $\pm$   0.095   &       -0.592  $\pm$   0.070   &       -0.617  $\pm$   0.063   \\
        &       5511.0-5521.0 & 0.401   $\pm$   0.150   &       0.221   $\pm$   0.127   &       0.145   $\pm$   0.110   \\
        &       6010.0-6030.0 & 0.298   $\pm$   0.067   &       0.042   $\pm$   0.065   &       -0.103  $\pm$   0.061   \\
        
  &  &   &  & \\
  
[Ni/Fe] &       4600.0-4610.0 & -0.253  $\pm$   0.236   &       -0.337  $\pm$   0.234   &       -0.430  $\pm$   0.224   \\
        &       4644.0-4654.0 & 0.431   $\pm$   0.175   &       0.423   $\pm$   0.176   &       0.311   $\pm$   0.163   \\
        &       4681.0-4691.0 & -0.548  $\pm$   0.346   &       -0.343  $\pm$   0.267   &       -0.277  $\pm$   0.226   \\
        &       4709.0-4719.0 & -0.197  $\pm$   0.173   &       -0.489  $\pm$   0.192   &       -0.592  $\pm$   0.188   \\
        &       4824.0-4835.0 & -1.302  $\pm$   0.247   &       -1.232  $\pm$   0.197   &       -1.242  $\pm$   0.179   \\
        &       4899.0-4909.0 & -0.588  $\pm$   0.221   &       -0.647  $\pm$   0.193   &       -0.643  $\pm$   0.179   \\
        &       4975.0-4985.0 & 0.470   $\pm$   0.102   &       0.471   $\pm$   0.097   &       0.402   $\pm$   0.089   \\
        &       5141.0-5151.0 & 0.412   $\pm$   0.158   &       0.109   $\pm$   0.149   &       0.047   $\pm$   0.154   \\
        &       5472.0-5482.0 & 0.317   $\pm$   0.037   &       0.255   $\pm$   0.035   &       0.128   $\pm$   0.037   \\
        &       5707.0-5717.0 & 1.215   $\pm$   0.075   &       0.849   $\pm$   0.088   &       0.642   $\pm$   0.089   \\
        &       6103.0-6113.0 & 0.689   $\pm$   0.069   &       0.521   $\pm$   0.064   &       0.456   $\pm$   0.061   \\
        &       6172.0-6182.0 & 0.151   $\pm$   0.055   &       0.055   $\pm$   0.049   &       0.011   $\pm$   0.044   \\
        
  &  &   &  & \\
  
[Ba/Fe] &       4551.0-4560.0 & 0.559   $\pm$   0.253   &       0.615   $\pm$   0.231   &       0.880   $\pm$   0.216   \\
        &       5849.0-5859.0 & 1.981   $\pm$   0.115   &       2.014   $\pm$   0.151   &       1.772   $\pm$   0.161   \\
        &       6135.0-6145.0 & 1.175   $\pm$   0.041   &       1.226   $\pm$   0.040   &       1.283   $\pm$   0.040   \\
        &       6492.0-6502.0 & 0.801   $\pm$   0.229   &       0.843   $\pm$   0.233   &       0.699   $\pm$   0.228   \\

  &  &   &  & \\
  
[Si/Fe] &       4420.5-4430.5   &       0.811   $\pm$   0.115   &       -0.452  $\pm$   0.494   &       0.150   $\pm$   0.120   \\
        &       4596.0-4606.0 & 0.027   $\pm$   0.310   &       1.975   $\pm$   0.096   &       1.151   $\pm$   0.095   \\
        &       5767.0-5777.0 & 0.775   $\pm$   0.227   &       0.118   $\pm$   0.221   &       0.413   $\pm$   0.133   \\
        &       5944.0-5954.0 & 0.277   $\pm$   0.080   &       0.940   $\pm$   0.231   &       0.750   $\pm$   0.281   \\
        &       6150.0-6160.0 & -0.476  $\pm$   0.073   &       -0.384  $\pm$   0.069   &       -0.322  $\pm$   0.065   \\
        &       6233.0-6242.0 & -0.373  $\pm$   0.102   &       -0.270  $\pm$   0.093   &       -0.227  $\pm$   0.088   \\
        &       7400.0-7427.0 & 0.078   $\pm$   0.027   &       0.116   $\pm$   0.028   &       0.154   $\pm$   0.028   \\

  &  &   &  & \\
  
[Cu/Fe] &       5101.0-5112.0 & 0.215   $\pm$   0.172   &       -0.033  $\pm$   0.189   &       -0.165  $\pm$   0.179   \\

\end{longtable}
\end{document}